\documentclass[12pt]{article}
\usepackage{draft}
\usepackage{cite}
\usepackage{cancel}
\usepackage{mathrsfs}
\usepackage{enumitem}
\usepackage{xcolor}
\usepackage{caption}  
\usepackage{graphicx} 
\usepackage{float} 
\usepackage{cite}
\usepackage{comment}
\usepackage{relsize}
\usepackage{physics}
\usepackage{psfrag}
\usepackage{cancel}
\usepackage{array}
\usepackage{amssymb}
\usepackage{amsmath}
\usepackage[compat=1.1.0]{tikz-feynman}
\usepackage{amsthm}
\usepackage{float}
\usepackage{tikz}
\usetikzlibrary{patterns}
\usetikzlibrary{decorations.pathmorphing}
\usetikzlibrary{mindmap,decorations.pathmorphing,backgrounds,positioning,fit}
\usetikzlibrary{decorations.markings}
\usepackage{tikz,lipsum,lmodern}
\usepackage[most]{tcolorbox}
\usepackage{hyperref}
\usepackage{xcolor}
\usepackage{tikz}
\usetikzlibrary{decorations.pathmorphing,patterns}
\usepackage{amsmath}
\usepackage{comment}
\usepackage{amssymb}
\usepackage{float}
\usepackage{fixmath}
\usepackage{physics}
\usepackage{slashed}
\usepackage{graphicx}
\usepackage{mathrsfs}
\usepackage{amsbsy}
\usepackage{subfig}
\usepackage{multirow}
\usepackage{hyperref}
\usepackage{bbm}
\usepackage{physics}
\usepackage{amsmath}
\usepackage{tikz}
\usepackage{mathdots}
\usepackage{yhmath}
\usepackage{cancel}
\usepackage{color}
\usepackage{siunitx}
\usepackage{array}
\usepackage{multirow}
\usepackage{amssymb}
\usepackage{gensymb}
\usepackage{tabularx}
\usepackage{extarrows}
\usepackage{booktabs}
\usetikzlibrary{fadings}
\usetikzlibrary{patterns}
\usetikzlibrary{shadows.blur}
\usetikzlibrary{shapes}
\usetikzlibrary{decorations.markings}
\tikzset{mid arrow/.style={postaction={decorate}, decoration={markings, mark=at position 0.5 with {\arrow{Triangle[angle=45:.21cm 1.5]}}}}}
\tikzset{
pattern size/.store in=\mcSize, 
pattern size = 5pt,
pattern thickness/.store in=\mcThickness, 
pattern thickness = 0.3pt,
pattern radius/.store in=\mcRadius, 
pattern radius = 1pt}
\makeatletter
\pgfutil@ifundefined{pgf@pattern@name@_cu8hrzqwm}{
\pgfdeclarepatternformonly[\mcThickness,\mcSize]{_cu8hrzqwm}
{\pgfqpoint{0pt}{0pt}}
{\pgfpoint{\mcSize+\mcThickness}{\mcSize+\mcThickness}}
{\pgfpoint{\mcSize}{\mcSize}}
{
\pgfsetcolor{\tikz@pattern@color}
\pgfsetlinewidth{\mcThickness}
\pgfpathmoveto{\pgfqpoint{0pt}{0pt}}
\pgfpathlineto{\pgfpoint{\mcSize+\mcThickness}{\mcSize+\mcThickness}}
\pgfusepath{stroke}
}}
\makeatother
\definecolor{LightGray}{rgb}{0.8, 0.8, 0.8}
\definecolor{twilightlavender}{rgb}{0.54, 0.29, 0.42}
\definecolor{richmaroon}{rgb}{0.69, 0.19, 0.38}
\definecolor{forestgreen(web)}{rgb}{0.13, 0.55, 0.13}
\definecolor{lava}{rgb}{0.81, 0.06, 0.13}
\hypersetup{
	breaklinks,
	colorlinks,
	citecolor=forestgreen(web),
	filecolor=richmaroon,
	linkcolor=lava,
	urlcolor=twilightlavender
}

\usepackage{cleveref}

\crefformat{section}{\S#2#1#3} 
\crefformat{subsection}{\S#2#1#3}
\crefformat{subsubsection}{\S#2#1#3}

\usepackage{verbatim}
 
\usepackage{color}

\newcommand{\out}{\text{out}}
\newcommand{\inm}{\text{in}}

\preprint{TIFR/TH/25-21}\title{Magnetic corrections to the classical soft photon theorems at all orders}
\affiliation[a]{Yau Mathematical Sciences Center (YMSC), Tsinghua University, Beijing 100084, China}
\affiliation[b]{Department of Theoretical Physics (DTP),
Tata Institute of Fundamental Research (TIFR), Homi Bhabha Road, Mumbai 400005, India}
\usepackage{orcidlink}
\author[a,\orcidlink{0000-0002-4535-3198}]{Sarthak Duary}\emailAdd{sarthakduary@tsinghua.edu.cn}
\author[b,\orcidlink{0000-0002-5553-7003}]{and Pabitra Ray}\emailAdd{raypabitra96@gmail.com}

\abstract{When a set of charged or dyonic objects scatter and subsequently disperse, the process generically emits electromagnetic radiation. The classical soft photon theorem constrains the constant term and leading power-law fall-off of the emitted waveform at asymptotic times solely in terms of the momenta and charges of the incoming and outgoing particles. In this work, we extend the analysis to include the full tower of subleading corrections in the late-time and early-time expansion of the electric and magnetic waveforms, providing explicit expressions that depend only on the kinematic data and the electric and magnetic charges of the scattered objects. Using independent electric and magnetic potentials, we compute the full classical soft expansion for general dyonic scattering. We provide explicit, recursive expressions for the trajectory coefficients and the resulting electric and magnetic waveforms. The resulting soft expansion exhibits manifest covariance under the electric--magnetic duality. The electric and magnetic waveforms transform as an $SO(2)$ doublet under electric--magnetic duality. This provides a non-trivial consistency check on our formulation. For the specific case of two-body scattering, we obtain resummed expressions for the waveforms in both the time and frequency domains.}
\begin{document}
\maketitle

\section{Introduction}
Soft photon theorems describe one of the most universal aspects of electrodynamics: the infrared behavior of radiation emitted during the scattering of charged particles. In any process involving accelerated charges, long-wavelength photons are radiated, and the amplitudes for emitting such soft photons are completely determined by the charges and momenta of the external states, independent of the microscopic details of the interaction. These soft theorems, along with their associated memory effects and an infinite-dimensional symmetry group acting at null infinity, constitute what is now widely known as the \textit{infrared triangle} \cite{Strominger:2013jfa,He:2014laa,He:2014cra,Strominger:2014pwa,Campiglia:2015qka,Kapec:2015ena,Strominger:2017zoo}. This framework reveals profound connections among the infrared structure of massless fields, the asymptotic symmetries of flat spacetime, and the conservation laws that govern physical observables. It has emerged as a foundational pillar for exploring the intricate relationship between classical and quantum dynamics in the infrared regime of both gravity and gauge theories.\vspace{0.4cm}\\
In \cite{Strominger:2015bla}, it was shown that the standard expression for the soft factor, which characterizes how scattering amplitudes behave when a photon with very low energy is emitted, must be modified in the presence of magnetic monopoles. The usual soft photon theorem connects scattering amplitudes with and without an additional soft photon and involves a soft factor that becomes singular as the photon’s momentum approaches zero. This relation holds exactly when there are no magnetic monopoles among the asymptotic particles. However, it was demonstrated that when monopoles are present, the general structure of the relation remains valid, but the explicit form of the soft factor changes. By employing electric--magnetic duality, a corrected expression was obtained. This modification plays an important role in understanding the infrared behavior of abelian gauge theories that include magnetic monopoles.\vspace{0.4cm}\\
When a collection of charged or dyonic objects undergoes a scattering event and subsequently disperses to infinity, the dynamics inevitably produce classical radiation. The classical soft photon theorem fixes the constant term and the leading power-law decay of the emitted radiation waveform at asymptotically early and late retarded times purely in terms of the asymptotic momenta and charges of the scattered particles. In this work, we generalize this result by computing the entire tower of subleading corrections in both the late- and early-time expansions of the electric and magnetic waveforms. This gives magnetic corrections to the classical soft photon theorem at all orders, with the resulting soft expansion manifestly covariant under electric--magnetic duality, offering a nontrivial consistency check of the formalism. For the specific case of two-body scattering, we further present resummed closed-form expressions for the complete low-frequency electric and magnetic waveforms in both time and frequency domains.\vspace{0.4cm}\\
The universality of the soft structure of electromagnetic radiation has deep roots in both classical dynamics and quantum scattering theory. In a series of foundational works \cite{Saha:2019tub, Laddha:2018myi, Laddha:2018vbn, Sahoo:2018lxl}, it was shown that the electromagnetic waveform observed at future null infinity (\(\mathscr{I}^+\)) exhibits a universal late- and early-time behavior: the coefficients of the constant term, the \(1/u\) term, and the \(u^{-2} \ln|u|\) term (with \(u\) the retarded time) are fully determined by the incoming and outgoing momenta and charges. Recently, this universality was extended in \cite{Karan:2025ndk, Duary:2025siq}, where it was demonstrated that all terms of the form \(u^{-r-1}(\ln|u|)^r\) for \(r \geq 0\) in the waveform are likewise fixed entirely by asymptotic kinematic and charge data-without reference to the details of the short-distance interaction. See also \cite{Alessio:2024onn, Fucito:2024wlg, Sen:2024bax, Georgoudis:2025vkk} for related studies.\vspace{0.4cm}\\
There is a correspondence between quantum soft theorems for \(\mathcal{S}\)-matrix elements and the asymptotic structure of classical radiation. For instance, the electromagnetic waveform at orders \(u^{-1}\) and \(u^{-2} \ln|u|\) has been shown to match the classical limit of quantum soft photon theorems at orders \(\ln \omega\) and \(\omega (\ln \omega)^2\), respectively, where \(\omega\) is the soft photon energy. This link was derived in \cite{Sahoo:2018lxl} by analyzing loop-level QED \(\mathcal{S}\)-matrices. More generally, an analysis of \((r+1)\)-loop QED amplitudes is expected to produce a soft photon theorem of order \(\omega^r (\ln \omega)^{r+1}\), whose classical limit precisely reproduces the electromagnetic waveform of order \(u^{-r-1} (\ln u)^r\) derived in \cite{Karan:2025ndk}.\vspace{0.4cm}\\Building on this quantum–classical correspondence, our work completes the classical picture by incorporating magnetic charges and extending the soft expansion to all orders in a duality-covariant manner. The inclusion of dyons not only enriches the physical content-allowing for the most general long-range electromagnetic interaction-but also imposes stringent constraints through electric--magnetic duality, which our results satisfy manifestly. By providing explicit, all-orders expressions for the electric and magnetic waveforms, we establish a fully universal description of soft classical radiation in generic dyonic scattering.\vspace{0.4cm}\\
In subsequent developments \cite{Strominger:2013lka, He:2015zea}, it was realized that, in the absence of monopoles, the conventional soft photon theorem can be interpreted as a Ward identity associated with an infinite-dimensional group of asymptotic symmetries-namely, large abelian gauge transformations that do not vanish at infinity. These symmetries lead to infinitely many conservation laws, relating moments of the electric field at past and future null infinity. The analysis in \cite{Strominger:2015bla} extended this framework to the magnetic sector. Abelian gauge theories also possess magnetic gauge symmetries that act on the dual magnetic potentials. When magnetic charges are included, there exists an infinite-dimensional subgroup of these symmetries that acts nontrivially on the $\mathcal{S}$-matrix. These magnetic gauge symmetries naturally combine with their electric counterparts into a complexified symmetry group. Within this setting, the modified soft photon theorem corresponds to the Ward identity of complexified large gauge transformations. All these symmetries are spontaneously broken, with the soft photons serving as the associated Goldstone bosons. Conversely, the corrected soft theorem also implies the existence of an infinite number of conserved electric and magnetic charges together with their corresponding symmetries. Finally, conservation of electric and magnetic charges ensures that the corrected soft factor remains invariant under both electric and magnetic gauge transformations. The magnetically corrected soft photon theorem can be understood as the Ward identity of complexified large electromagnetic gauge symmetries. \vspace{0.2cm}\\
\subsection*{\colorbox{LightGray}{Main results.}}
In this work, we analyze the complete tower of soft corrections in the electromagnetic waveform emitted by general dyonic scattering. We adopt the local two-potential formulation \cite{Rohrlich:1966zz, Zwanziger:1970hk, Schwinger:1966nj, Zwanziger:1968rs, Yan1966, Zwanziger:1968ams, Edwards:2022dbd , Blagojevic:1979bm, Blagojevic:1985sh, Blagojevic:1981he, Blagojevic:1982bv}. See \cite{Choi:2019sjs, Csaki:2020inw, Csaki:2020yei, Csaki:2021ozp, Csaki:2022tvb} for recent works. We show that the classical expansion exhibits manifest covariance under electric--magnetic duality and that all coefficients in the expansion can be expressed purely in terms of the kinematic data and the electric and magnetic charges of the scattered objects. In particular, we extend the classical soft photon theorem to include magnetic charges, providing explicit all-orders expressions for the electric and magnetic waveforms emitted during generic dyonic scattering. The resulting soft expansion includes the full tower of subleading terms of the form $u^{-r-1}(\ln|u|)^r$ ($r\geq0$) at both early and late retarded times, determined solely by the asymptotic kinematic data and the electric and magnetic charges of the scattered particles. A central feature of the result is its manifest covariance under electric--magnetic duality. Specifically,
\begin{quotation}
\noindent\textit{The electric and magnetic waveforms transform as an $SO(2)$ doublet under electric--magnetic duality.}
\end{quotation}
That is, under a duality rotation by angle $\theta$, \begin{equation}
\begin{pmatrix}
A_e^\mu \\
A_m^\mu
\end{pmatrix}
\;\longrightarrow\;
\begin{pmatrix}
\cos\theta & \sin\theta \\
-\sin\theta & \phantom{-}\cos\theta
\end{pmatrix}
\begin{pmatrix}
A_e^\mu \\
A_m^\mu
\end{pmatrix}~.
\end{equation}
The covariance of the full waveform structure under this symmetry provides a nontrivial consistency check of the formalism and completes the classical picture of soft electromagnetic radiation in the presence of magnetic monopoles.

\subsection*{\colorbox{LightGray}{Set up.}}

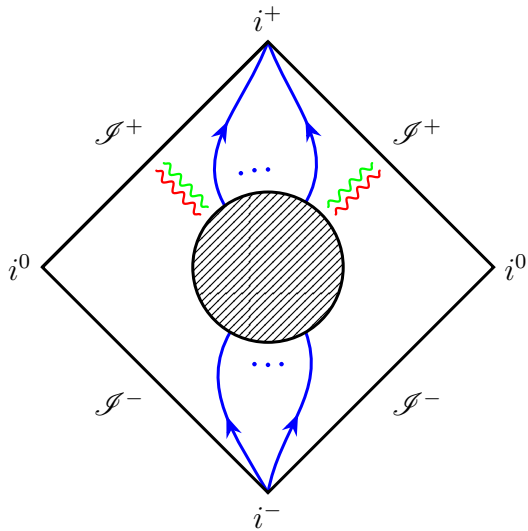
\begin{figure}[h]
\begin{center}
\begin{tikzpicture}[mid arrow/.style={postaction={decorate},decoration={markings,mark=at position 0.5 with {\arrow{Stealth}}}}]

\draw[mid arrow,line width=0.4mm,blue] (-0.57,0.81) to [out=120,in=-110] (0,3);
\draw[mid arrow,line width=0.4mm,blue] (0.5,0.86) to [out=60,in=-70] (0,3);
\draw[mid arrow,line width=0.4mm,blue] (0,-3) to [out=80,in=-70] (0.5,-0.86);
\draw[mid arrow,line width=0.4mm,blue] (0,-3) to [out=120,in=-120] (-0.5,-0.86);

\fill[blue] (0,1.3) circle [radius=0.03cm];
\fill[blue] (-0.18,1.29) circle [radius=0.03cm];
\fill[blue] (-0.36,1.25) circle [radius=0.03cm];

\fill[blue] (-0.18,-1.29) circle [radius=0.03cm];
\fill[blue] (0,-1.3) circle [radius=0.03cm];
\fill[blue] (0.18,-1.29) circle [radius=0.03cm];

\begin{scope}[xshift=1mm, yshift=-1mm, rotate=0]
 \draw[line width=0.3mm,decorate, decoration={snake, amplitude=0.5mm, segment length=2mm},color=red,line cap=round]  (0.8,0.8) -- (1.385,1.385); 
\end{scope}

\begin{scope}[xshift=-1mm, yshift=-1mm, rotate=0]
\draw[line width=0.3mm,decorate, decoration={snake, amplitude=0.5mm, segment length=2mm},color=red,line cap=round]  (-0.8,0.8) -- (-1.385,1.385);  
\end{scope}

\node[above right] at (1.5,1.5) {$\mathscr{I}^+$};
\node[above left] at (-1.5,1.5) {$\mathscr{I}^+$};
\node[below right] at (1.5,-1.5) {$\mathscr{I}^-$};
\node[below left] at (-1.5,-1.5) {$\mathscr{I}^-$};

\node[right] at (3,0) {$i^0$};
\node[left] at (-3,0) {$i^0$};
\node[above] at (0,3) {$i^+$};
\node[below] at (0,-3) {$i^-$};

\draw[line width=0.3mm,decorate, decoration={snake, amplitude=0.5mm, segment length=2mm},color=green,line cap=round]  (0.8,0.8) -- (1.385,1.385);
\draw[line width=0.3mm,decorate, decoration={snake, amplitude=0.5mm, segment length=2mm},color=green,line cap=round]  (-0.8,0.8) -- (-1.385,1.385);

\draw[line width=0.4mm] (0,3) -- (-3,0) -- (0,-3) -- (3,0) -- (0,3) -- (-3,0);

\draw[pattern=north east lines,line width=0.4mm] (0,0) circle [radius=1cm];

\end{tikzpicture}
\caption{Scattering of dyons: incoming and outgoing dyon trajectories are shown in blue. Electric waveform is shown in green wiggly lines and the magnetic waveform is shown in red wiggly lines.}
\label{fig:dyon_scattering}
\end{center}
\end{figure}

We consider a scattering process involving \(N'\) incoming particles - now understood as dyons - each carrying a mass \(m'_a\), four-momentum \(p'_a\), and a pair of electromagnetic charges \((q'_a, g'_a)\). These particles undergo a dynamical interaction - possibly including collision, fusion, and subsequent fragmentation - resulting in \(N\) outgoing dyons, characterized by masses \(m_a\), momenta \(p_a\), and dyonic charges \((q_a, g_a)\). 
The electric and magnetic waveforms propagate outward and are asymptotically detected at future null infinity, denoted \(\mathscr{I}^+\). Let \(\vb{x}\) denote the spatial position of a distant detector located at \(\mathscr{I}^+\), where we measure the asymptotic electric and magnetic waveforms generated by the scattering event. The setup is shown in \figurename\,[\ref{fig:dyon_scattering}].
To describe the geometry of detection, we define the following quantities
\begin{equation}\label{eq:conventions}
R \equiv \abs{\vb{x}}~, \quad \hat{\vb{n}} \equiv \frac{\vb{x}}{R}~, \quad n \equiv (1, \hat{\vb{n}})~, \quad u \equiv x^0 - R~,
\end{equation}
where \(u\) is the retarded time and we work in natural units (\(c = 1\)). The peak of the electric and magnetic waveforms typically arrives at the detector around \(u \sim 1\). Our primary interest lies in the asymptotic behavior of the electric and magnetic waveforms as \(u \to +\infty\), which is determined by the late-time trajectories of the outgoing dyons. For the \(a\)-th outgoing dyon, the relevant portion of its worldline trajectory \cite{Karan:2025ndk,Alessio:2024onn} is given by 
\begin{equation}\label{eq:trajectory}
X^\mu_a(\tau_a) = \frac{p^\mu_a}{m_a} \tau_a + \left( C^{(a)\mu}_1 \ln\tau_a + \mathcal{O}(1) \right) - \sum_{r=1}^{\infty} \left( C^{(a)\mu}_{r+1} \left( \frac{\ln\tau_a}{\tau_a} \right)^r + \mathcal{O}\big( \tau_a^{-r} (\ln\tau_a)^{r-1} \big) \right) + \cdots~~.
\end{equation}
Here, \(\tau_a\) is a parameter along the particle’s worldline (e.g., proper time), \(p^\mu_a\) and \(m_a\) are its momentum and mass, and the coefficients \(C^{(a)\mu}_{r+1}\) encode corrections due to electromagnetic self-interaction - now sensitive to both electric charge \(q_a\) and magnetic charge \(g_a\), reflecting the duality-symmetric nature of the underlying theory.\vspace{0.4cm}\\
These trajectory corrections directly influence the structure of the asymptotic electromagnetic field. In particular, they generate terms in the electric and magnetic waveforms that scale as
\begin{equation}
u^{-r-1} (\ln |u|)^r \quad \text{for all } r \geq 0~.
\end{equation}
Our goal is to preserve all such subleading logarithmic terms in the expansion of the dyonic electromagnetic waveform. To facilitate this analysis, we introduce a large formal parameter \(\lambda\) and assume that both \(u\) and \(\ln u\) scale as \(\lambda\). When computing the electric and magnetic waveforms measured at retarded time \(u\), the corresponding worldline parameter \(\tau_a\) for the \(a\)-th outgoing dyon is also of order \(u\); hence \(\tau_a \sim \lambda\) and \(\ln \tau_a \sim \ln u\).\vspace{0.4cm}\\
For the incoming dyons, we analytically continue the trajectory to the past by replacing \(\tau_a\) with \(-\tau_a\) (with \(\tau_a > 0\)), and \(\ln \tau_a\) with \(\ln |\tau_a|\). This substitution introduces an additional factor of \((-1)^r\) in the coefficient \(C^{(a)\mu}_{r+1}\) at each order \(r\). Simultaneously, all outgoing quantities - momenta, masses, and dyonic charges \((q_a, g_a)\) - are replaced by their incoming counterparts \((p'_a, m'_a, q'_a, g'_a)\).\vspace{0.4cm}\\
This framework allows us to systematically compute the full asymptotic structure of the electric–magnetic radiation field produced in dyonic scattering, including memory effects, tail terms, and higher-order logarithmic corrections - all of which depend nontrivially on the dyonic charge content of the scattered particles.

\section{Dual electromagnetic fields and electric--magnetic duality}

This section introduces a duality-symmetric formulation of electrodynamics capable of describing dyons-particles carrying both electric and magnetic charges. Standard electrodynamics, based on a single gauge potential \(A_\mu\), enforces the Bianchi identity \(\partial_{[\lambda}F_{\mu\nu]} = 0\), which excludes magnetic monopoles. To accommodate both types of charge, the theory is extended by introducing two independent four-potentials: an electric potential \(A_e^\mu\) and a magnetic potential \(A_m^\mu\). 
To describe dyons---particles carrying both electric and magnetic charges---we must modify this formulation so that both $F^{\mu\nu}$ and its dual $\widetilde{F}^{\mu\nu}$ can act as independent sources. A manifestly symmetric description introduces two vector potentials
$$
A^\mu_e \quad \text{(electric potential)}~, \qquad A^\mu_m \quad \text{(magnetic potential)}~.
$$
These fields together encode the electromagnetic field tensor and its dual in a way that treats the electric and magnetic sectors on equal footing.
We define the generalized field strength by
\begin{equation}
\begin{split}
F^{\mu\nu} 
&\equiv \partial^\mu A^\nu_e - \partial^\nu A^\mu_e 
 - \epsilon^{\mu\nu\rho\sigma} \partial_\rho A_{m\,\sigma}~,\\[4pt]
\widetilde{F}^{\mu\nu} 
&\equiv \partial^\mu A^\nu_m - \partial^\nu A^\mu_m 
 + \epsilon^{\mu\nu\rho\sigma} \partial_\rho A_{e\,\sigma}~.
\end{split}
\end{equation}
The two tensors are related by 
\begin{equation}
\widetilde{F}^{\mu\nu} = \tfrac{1}{2} \epsilon^{\mu\nu\rho\sigma} F_{\rho\sigma}~.
\end{equation}
Maxwell equations in the presence of both electric and magnetic currents\footnote{
Both currents are conserved. } $J_e^\mu$ and $J_m^\mu$ are given by
\begin{equation}\label{eq:DualMaxwellEquation}
\begin{split}
\partial_\nu F^{\nu\mu} &= - J_e^\mu~,\\
\partial_\nu \widetilde{F}^{\nu\mu} &= - J_m^\mu~.
\end{split}
\end{equation}
\subsection{Electric--magnetic duality}
The theory defined by the equations of motion \eqref{eq:DualMaxwellEquation} is invariant under continuous rotations mixing the electric and magnetic fields
\begin{equation}
\begin{split}
\begin{pmatrix}
F' \\[2pt] \widetilde{F}'
\end{pmatrix}
&=
\begin{pmatrix}
\cos\theta & \sin\theta \\[2pt]
-\sin\theta & \cos\theta
\end{pmatrix}
\begin{pmatrix}
F \\[2pt] \widetilde{F}
\end{pmatrix}~.
\end{split}
\end{equation}
This is the \emph{electric--magnetic duality} symmetry, an internal $SO(2)$ rotation in the space of field strengths. \vspace{0.4cm}\\To maintain invariance of Maxwell’s equations, the sources must transform in the same way
\begin{equation}
\begin{split}
\begin{pmatrix}
J_e' \\[2pt] J_m'
\end{pmatrix}
&=
\begin{pmatrix}
\cos\theta & \sin\theta \\[2pt]
-\sin\theta & \cos\theta
\end{pmatrix}
\begin{pmatrix}
J_e \\[2pt] J_m
\end{pmatrix}~.
\end{split}
\end{equation}
Correspondingly, the potentials transform as a $SO(2)$ doublet
\begin{equation}
\begin{split}
\begin{pmatrix}
A_e' \\[2pt] A_m'
\end{pmatrix}
&=
\begin{pmatrix}
\cos\theta & \sin\theta \\[2pt]
-\sin\theta & \cos\theta
\end{pmatrix}
\begin{pmatrix}
A_e \\[2pt] A_m
\end{pmatrix}~.
\label{eq:SO2rotation}
\end{split}
\end{equation}
\subsection*{\colorbox{LightGray}{Physical interpretation.}}
The $SO(2)$ duality symmetry expresses the \emph{indistinguishability} of the electric and magnetic sectors at the level of the free Maxwell equations. Consider a point particle\footnote{The worldline formulation of the theory for a point particle having both electric and magnetic charges $(q,g)$ can be found in \cite{Rohrlich:1966zz,Zwanziger:1970hk}.}(dyon) of mass $m$ carrying both electric and magnetic charges $(q,g)$. The existence of a dyon implies that nature may realize a \emph{rotational symmetry} between charge types, where electric and magnetic charges transform as
    \begin{equation}
    \begin{split}
    \begin{pmatrix}
    q' \\ g'
    \end{pmatrix}
    &=
    \begin{pmatrix}
    \cos\theta & \sin\theta \\[2pt]
    -\sin\theta & \cos\theta
    \end{pmatrix}
    \begin{pmatrix}
    q \\ g
    \end{pmatrix}~.
    \end{split}
    \end{equation}
Quantum mechanically, this $SO(2)$ duality is restricted to a discrete subgroup $SL(2,\mathbb{Z})$ due to Dirac-Schwinger-Zwanziger charge quantization. Thus, the dual-symmetric Maxwell theory not only unifies electric and magnetic fields geometrically but also reveals an underlying rotational symmetry in the charge--field space that becomes central in supersymmetric and string theories, where electric--magnetic duality extends to full \emph{S-duality}.

\section{Trajectory coefficients}
The equations of motion that govern the dynamics of the trajectory of the multiple dyons $X^\mu_a(\tau_a)$ and the vector potentials $A^\mu_{e\,(b)}(x)$ and $A^\mu_{m\,(b)}(x)$ are given by
\begin{equation}
\label{eq:equations_of_motion}
 \begin{split}
&m_a\frac{d^2X^\mu_a(\tau_a)}{d\tau_a^2}=\frac{dX_{a\nu}(\tau_a)}{d\tau_a}\sum_{b\neq a}\Big[q_aF^{\mu\nu}_{(b)}(X_a(\tau_a))+g_a\widetilde{F}^{\mu\nu}_{(b)}(X_a(\tau_a))\Big]~,\\
&\square A^\mu_{e\,(b)}(x)=-J^\mu_{e\,(b)}(x)~,\quad \square A^\mu_{m\,(b)}(x)=-J^\mu_{m\,(b)}(x)~.
 \end{split}   
\end{equation}
The first equation is the generalization of the Lorentz force law in the presence of the magnetic monopole. The equations obeyed by the electric and magnetic vector potentials are Maxwell's equations written in the Lorenz gauge\footnote{In this gauge, they satisfy the following conditions 
\begin{equation}
\partial_\mu A^\mu_{e\,(b)}(x)=0~,\qquad\partial_\mu A^\mu_{m\,(b)}(x)=0~.    
\end{equation}}.
Currents $J^\mu_{e\,(b)}(x)$ and $J^\mu_{m\,(b)}(x)$ can be expressed in terms of charges \(q_b\) and \(g_b\) of the particle \( b \) as
\begin{equation}\label{eq:currents}
\begin{split}
J^\mu_{e\,(b)}(x)&= q_b \int_{0}^{\infty} d\tau_b\,\delta^{(4)}\big(x - X_b(\tau_b)\big)\frac{dX^\mu_b(\tau_b)}{d\tau_b}~,\\J^\mu_{m\,(b)}(x)&= g_b \int_{0}^{\infty} d\tau_b\,\delta^{(4)}\big(x - X_b(\tau_b)\big) \frac{dX^\mu_b(\tau_b)}{d\tau_b} ~.
\end{split}
\end{equation}
The retarded Green function solutions for $A^\mu_{e\,(b)}(x)$ and $A^\mu_{m\,(b)}(x)$ are given by
\begin{equation}
\begin{split}
A^\mu_{e\,(b)}(x)&=\frac{q_b}{2\pi}\int_{0}^{\infty} d\tau_b~\delta_+\Big(-\big(x-X_b(\tau_b)\big)^2\Big)\frac{dX^\mu_b(\tau_b)}{d\tau_b}~,\\A^\mu_{m\,(b)}(x)&=\frac{g_b}{2\pi}\int_{0}^{\infty} d\tau_b~\delta_+\Big(-\big(x-X_b(\tau_b)\big)^2\Big)\frac{dX^\mu_b(\tau_b)}{d\tau_b}~.    
\end{split}
\end{equation}
To calculate the derivative of the delta function, we have used the following identity
\begin{equation}
{\small
\begin{split}
\pdv{x_\mu}\delta_+\Big(-\big(x-X_b(\tau_b)\big)^2\Big)&=-2\big(x^\mu-X^\mu_{b}(\tau_b)\big)\delta'_+\Big(-\big(x-X_b(\tau_b)\big)^2\Big)\\&=-\frac{\big(x^\mu-X^\mu_{b}(\tau_b)\big)}{\big(x-X_{b}(\tau_b)\big)\cdot\frac{dX_b(\tau_b)}{d\tau_b}}\left[\pdv{\tau_b}\delta_+\Big(-\big(x-X_b(\tau_b)\big)^2\Big) \right]~. 
\end{split}}
\end{equation}
So, the field strength $F^{\mu\nu}_{(b)}(x)$ is given by
\begin{equation}
{\small
\begin{split}
&F^{\mu\nu}_{(b)}(x)\\&=\partial^\mu A^\nu_{e\,(b)}(x)-\partial^\nu A^\mu_{e\,(b)}(x)-\epsilon^{\mu\nu\rho\sigma}\partial_\rho A_{m\,(b)\,\sigma}(x)\\&=\frac{1}{2\pi}\int_{0}^{\infty} d\tau_b\,\delta_+\Big(-\big(x-X_b(\tau_b)\big)^2\Big)\\&~~\times \left[\frac{q_b\left\{\big(x^\mu-X_b^\mu(\tau_b)\big)\frac{d^2X^\nu_b(\tau_b)}{d\tau_b^2}-\big(x^\nu-X_b^\nu(\tau_b)\big)\frac{d^2X^\mu_b(\tau_b)}{d\tau_b^2}\right\}-g_b\epsilon^{\mu\nu\rho\sigma}\big(x_\rho-X_{b\rho}(\tau_b)\big)\frac{d^2X_{b\sigma}(\tau_b)}{d\tau_b^2}}{\big(x-X_b(\tau_b)\big)\cdot\frac{dX_b(\tau_b)}{d\tau_b}}\right.\\&~~~~~~-\frac{q_b\left\{\big(x^\mu-X_b^\mu(\tau_b)\big)\frac{dX^\nu_b(\tau_b)}{d\tau_b}-\big(x^\nu-X_b^\nu(\tau_b)\big)\frac{dX^\mu_b(\tau_b)}{d\tau_b}\right\}-g_b\epsilon^{\mu\nu\rho\sigma}\big(x_\rho-X_{b\rho}(\tau_b)\big)\frac{dX_{b\sigma}(\tau_b)}{d\tau_b}}{\left(\big(x-X_b(\tau_b)\big)\cdot\frac{dX_b(\tau_b)}{d\tau_b}\right)^2}\\&~~~~~~~~~~~~~~~~~~~~~~~~~~~~~~~~~~~~~~~~~~~~~~~~~\left.\vphantom{
\frac{g_b\left\{\big(x^\mu-X_b^\mu(\tau_b)\big)\frac{d^2X^\nu_b(\tau_b)}{d\tau_b^2}\right\}}{\big(x-X_b(\tau_b)\big)\cdot\frac{dX_b(\tau_b)}{d\tau_b}}}\times\left\{\big(x-X_b(\tau_b)\big)\cdot\frac{d^2X_b(\tau_b)}{d\tau_b^2}-\left(\frac{dX_b(\tau_b)}{d\tau_b}\right)^2\right\}\right] ~~.   
\end{split}  
}
\end{equation}
Similarly, the dual field strength $\widetilde{F}^{\mu\nu}_{(b)}(x)$ is given by
\begin{equation}
{\small
\begin{split}
&\widetilde{F}^{\mu\nu}_{(b)}(x)\\&=\partial^\mu A^\nu_{m\,(b)}(x)-\partial^\nu A^\mu_{m\,(b)}(x)+\epsilon^{\mu\nu\rho\sigma}\partial_\rho A_{e\,(b)\,\sigma}(x)\\&=\frac{1}{2\pi}\int_{0}^{\infty} d\tau_b\,\delta_+\Big(-\big(x-X_b(\tau_b)\big)^2\Big)\\&~~\times \left[\frac{g_b\left\{\big(x^\mu-X_b^\mu(\tau_b)\big)\frac{d^2X^\nu_b(\tau_b)}{d\tau_b^2}-\big(x^\nu-X_b^\nu(\tau_b)\big)\frac{d^2X^\mu_b(\tau_b)}{d\tau_b^2}\right\}+q_b\epsilon^{\mu\nu\rho\sigma}\big(x_\rho-X_{b\rho}(\tau_b)\big)\frac{d^2X_{b\sigma}(\tau_b)}{d\tau_b^2}}{\big(x-X_b(\tau_b)\big)\cdot\frac{dX_b(\tau_b)}{d\tau_b}}\right.\\&~~~~~~-\frac{g_b\left\{\big(x^\mu-X_b^\mu(\tau_b)\big)\frac{dX^\nu_b(\tau_b)}{d\tau_b}-\big(x^\nu-X_b^\nu(\tau_b)\big)\frac{dX^\mu_b(\tau_b)}{d\tau_b}\right\}+q_b\epsilon^{\mu\nu\rho\sigma}\big(x_\rho-X_{b\rho}(\tau_b)\big)\frac{dX_{b\sigma}(\tau_b)}{d\tau_b}}{\left(\big(x-X_b(\tau_b)\big)\cdot\frac{dX_b(\tau_b)}{d\tau_b}\right)^2}\\&~~~~~~~~~~~~~~~~~~~~~~~~~~~~~~~~~~~~~~~~~~~~~~~~\left.\vphantom{
\frac{g_b\left\{\big(x^\mu-X_b^\mu(\tau_b)\big)\frac{d^2X^\nu_b(\tau_b)}{d\tau_b^2}\right\}}{\big(x-X_b(\tau_b)\big)\cdot\frac{dX_b(\tau_b)}{d\tau_b}}}\times\left\{\big(x-X_b(\tau_b)\big)\cdot\frac{d^2X_b(\tau_b)}{d\tau_b^2}-\left(\frac{dX_b(\tau_b)}{d\tau_b}\right)^2\right\}\right]~. 
\end{split}}    
\end{equation}
To evaluate the right-hand side of the equations of motion for the particle \( a \), we must compute the electromagnetic field strength at its position, which entails evaluating \( F^{\mu\nu}_{(b)}(x) \) (or \( \widetilde{F}^{\mu\nu}_{(b)}(x) \)) generated by the particle \( b \). This requires expressing the worldline parameter \( \tau_b \) of the particle \( b \) as a function of the parameter \( \tau_a \) of the particle \( a \), a relation obtained by solving the appropriate constraint equation 
\begin{equation}
\big(X_a(\tau_a) - X_b(\tau_b)\big)^2 = 0~.
\end{equation}
This condition is enforced by the delta function, which restricts the interaction to lie on the light-cone. We now need to carefully determine which contributions from the trajectory expansions \( X^\mu_a(\tau_a) \) and \( X^\mu_b(\tau_b) \) are relevant and which may be safely discarded. Since our goal is to retain all terms in the waveform that scale as \( u^{-r-1} (\ln u)^r \), we introduce a bookkeeping parameter \( \lambda \) and adopt the scaling \( u \sim \ln u \sim \lambda \). This prescription allows us to consistently organize and compare the magnitudes of different terms in the expansion of the trajectories. A key observation is that, when evaluating the waveform at a detector located at retarded time \( u \), the relevant worldline parameters \( \tau_a \) and \( \tau_b \) are themselves of order \( u \). It follows that \( \tau_a \sim \lambda \) and \( \tau_b \sim \lambda \), and since \( \ln \tau_a \simeq \ln u \) and \( \ln \tau_b \simeq \ln u \), their logarithms also scale as \( \lambda \). Inspecting the expansion of \( X^\mu_a(\tau_a) \), we see that the terms linear in \( \tau_a \) and those involving \( \ln \tau_a \) are both \( \mathcal{O}(\lambda) \), whereas the coefficients \( C_r^{(a)\mu} \) for \( r \geq 2 \) remain \( \mathcal{O}(\lambda^0) \). Meanwhile, any discarded terms in the expansion of \( X^\mu_a(\tau_a) \) scale as \( \lambda^{-p} \) with \( p \geq 1 \), rendering them subdominant and irrelevant for the purposes of our calculation. Thus, in solving the light-cone condition \( \big( X_a(\tau_a) - X_b(\tau_b) \big)^2 = 0 \) to express \( \tau_b \) as a function of \( \tau_a \), we retain only the leading contributions from the trajectory expansions of \( X^\mu_a(\tau_a) \) and \( X^\mu_b(\tau_b) \). Concretely, we keep the first two terms in each expansion and neglect all higher-order corrections. To solve the constraint equation \(\big(X_a(\tau_a) - X_b(\tau_b)\big)^2 = 0\), we will employ the following expansion  
\begin{equation}\label{eq:trajactory_leading}
X^\mu_c(\tau_c)=\frac{p^\mu_c}{m_c}\tau_c+C^{(c)\mu}_1\ln\tau_c+\mathcal{O}(\lambda^0)\quad\textmd{where}~~c=a,b~~.
\end{equation}
Equation \(\big(X_a(\tau_a) - X_b(\tau_b)\big)^2 = 0\) simplifies to
\begin{equation}
\begin{split}
 &\tau_a^2+\tau_b^2+2\frac{p_a\cdot p_b}{m_am_b}\tau_a\tau_b+2\frac{p_a\cdot\left( C^{(b)}_1-C^{(a)}_1\right)}{m_a}\tau_a\ln\tau_a-2\frac{p_b\cdot\left( C^{(b)}_1-C^{(a)}_1\right)}{m_b}\tau_b\ln\tau_a\\&~~~~-\left( C^{(b)}_1-C^{(a)}_1\right)^2(\ln\tau_a)^2+\mathcal{O}(\lambda)=0~.
\end{split} 
\end{equation}
The solution for $\tau_b$ is 
\begin{equation}
\tau_b=\tau_{ba}^\star+\mathcal{O}(\lambda^0)~,
\end{equation}
and $\tau_{ba}^\star$ is expressed as 
\begin{equation}
\begin{split}
\tau_{ba}^\star&=-\frac{p_a\cdot p_b}{m_am_b}\tau_a+\frac{p_b\cdot\left(C_1^{(b)}-C^{(a)}_1\right)}{m_b}\ln\tau_a-\left[-\tau_a^2-2\frac{p_a\cdot\left(C_1^{(b)}-C^{(a)}_1\right)}{m_a}\tau_a\ln\tau_a\vphantom{\left(\frac{p_b\cdot\left(C_1^{(b)}-C^{(a)}_1\right)}{m_b}\right)^2}\right.\\&~~~~+\left. \left(C_1^{(b)}-C^{(a)}_1\right)^2(\ln\tau_a)^2+\left(\frac{p_a\cdot p_b}{m_am_b}\tau_a-\frac{p_b\cdot\left(C_1^{(b)}-C^{(a)}_1\right)}{m_b}\ln\tau_a\right)^2\right]^{\frac{1}{2}}~,
\end{split}   
\end{equation}
where we have selected the smallest value of \(\tau_b\), as dictated by the Heaviside step function \(\mathrm{\Theta} \big(X^0_a(\tau_a) - X^0_b(\tau_b)\big)\) appearing in the retarded Green function solution. Substituting $x^\mu=X^\mu_a(\tau_a)$ the delta function constraint gives us
\begin{equation}
\begin{split}
\delta_+\Big(-\big(X_a(\tau_a)-X_b(\tau_b)\big)^2\Big)=\frac{1}{2\abs{\big(X_a(\tau_a)-X_b(\tau_b)\big)\cdot\frac{dX_b(\tau_b)}{d\tau_b}}}_{\tau_b=\tau^\star_{ba}}\delta(\tau_b-\tau_{ba}^\star)~.
\end{split}    
\end{equation}
Evaluating the delta function integral, the final result for the field strength can be expressed as
\begin{equation}
{\footnotesize
\begin{split}
&F^{\mu\nu}_{(b)}(X_a(\tau_a))\\&=\frac{1}{4\pi\abs{\big(X_a(\tau_a)-X_b(\tau_b)\big)\cdot\frac{dX_b(\tau_b)}{d\tau_b}}}_{\tau_b=\tau^\star_{ba}}
\\&\times\left[\frac{q_b\left\{\big(X_a^\mu(\tau_a)-X_b^\mu(\tau_b)\big)\frac{d^2X^\nu_b(\tau_b)}{d\tau_b^2}-\big(X_a^\nu(\tau_a)-X_b^\nu(\tau_b)\big)\frac{d^2X^\mu_b(\tau_b)}{d\tau_b^2}\right\}-g_b\epsilon^{\mu\nu\rho\sigma}\big(X_{a\rho}(\tau_a)-X_{b\rho}(\tau_b)\big)\frac{d^2X_{b\sigma}(\tau_b)}{d\tau_b^2}}{\big(X_a(\tau_a)-X_b(\tau_b)\big)\cdot\frac{dX_b(\tau_b)}{d\tau_b}}\right.\\&~~-\frac{q_b\left\{\big(X_a^\mu(\tau_a)-X_b^\mu(\tau_b)\big)\frac{dX^\nu_b(\tau_b)}{d\tau_b}-\big(X_a^\nu(\tau_a)-X_b^\nu(\tau_b)\big)\frac{dX^\mu_b(\tau_b)}{d\tau_b}\right\}-g_b\epsilon^{\mu\nu\rho\sigma}\big(X_{a\rho}(\tau_a)-X_{b\rho}(\tau_b)\big)\frac{dX_{b\sigma}(\tau_b)}{d\tau_b}}{\left(\big(X_a(\tau_a)-X_b(\tau_b)\big)\cdot\frac{dX_b(\tau_b)}{d\tau_b}\right)^2}\\&~~~~~~~~~~~~~~~~~~~~~~~~~~~~~~~~~~~~~~~~~~~~~~~~~~~~~\left.\vphantom{\frac{q_b\left\{\big(X_a^\mu(\tau_a)-X_b^\mu(\tau_b)\big)\frac{d^2X^\nu_b(\tau_b)}{d\tau_b^2}-\big(X_a^\nu(\tau_a)-X_b^\nu(\tau_b)\big)\frac{d^2X^\mu_b(\tau_b)}{d\tau_b^2}\right\}+g_b\epsilon^{\mu\nu\rho\sigma}\big(X_{a\rho}(\tau_a)-X_{b\rho}(\tau_b)\big)\frac{d^2X_{b\sigma}(\tau_b)}{d\tau_b^2}}{\big(X_a(\tau_a)-X_b(\tau_b)\big)\cdot\frac{dX_b(\tau_b)}{d\tau_b}}}\times\left\{\big(X_a(\tau_a)-X_b(\tau_b)\big)\cdot\frac{d^2X_b(\tau_b)}{d\tau_b^2}-\left(\frac{dX_b(\tau_b)}{d\tau_b}\right)^2\right\}\right]_{\tau_b=\tau^\star_{ba}}~~,
\end{split}
}
\end{equation}
and the dual field strength is given by
\begin{equation}
{\footnotesize
\begin{split}
&\widetilde{F}^{\mu\nu}_{(b)}(X_a(\tau_a))\\&=\frac{1}{4\pi\abs{\big(X_a(\tau_a)-X_b(\tau_b)\big)\cdot\frac{dX_b(\tau_b)}{d\tau_b}}}_{\tau_b=\tau^\star_{ba}}
\\&\times\left[\frac{g_b\left\{\big(X_a^\mu(\tau_a)-X_b^\mu(\tau_b)\big)\frac{d^2X^\nu_b(\tau_b)}{d\tau_b^2}-\big(X_a^\nu(\tau_a)-X_b^\nu(\tau_b)\big)\frac{d^2X^\mu_b(\tau_b)}{d\tau_b^2}\right\}+q_b\epsilon^{\mu\nu\rho\sigma}\big(X_{a\rho}(\tau_a)-X_{b\rho}(\tau_b)\big)\frac{d^2X_{b\sigma}(\tau_b)}{d\tau_b^2}}{\big(X_a(\tau_a)-X_b(\tau_b)\big)\cdot\frac{dX_b(\tau_b)}{d\tau_b}}\right.\\&~~-\frac{g_b\left\{\big(X_a^\mu(\tau_a)-X_b^\mu(\tau_b)\big)\frac{dX^\nu_b(\tau_b)}{d\tau_b}-\big(X_a^\nu(\tau_a)-X_b^\nu(\tau_b)\big)\frac{dX^\mu_b(\tau_b)}{d\tau_b}\right\}+q_b\epsilon^{\mu\nu\rho\sigma}\big(X_{a\rho}(\tau_a)-X_{b\rho}(\tau_b)\big)\frac{dX_{b\sigma}(\tau_b)}{d\tau_b}}{\left(\big(X_a(\tau_a)-X_b(\tau_b)\big)\cdot\frac{dX_b(\tau_b)}{d\tau_b}\right)^2}\\&~~~~~~~~~~~~~~~~~~~~~~~~~~~~~~~~~~~~~~~~~~~~~~~~~~~~~\left.\vphantom{\frac{q_b\left\{\big(X_a^\mu(\tau_a)-X_b^\mu(\tau_b)\big)\frac{d^2X^\nu_b(\tau_b)}{d\tau_b^2}-\big(X_a^\nu(\tau_a)-X_b^\nu(\tau_b)\big)\frac{d^2X^\mu_b(\tau_b)}{d\tau_b^2}\right\}+g_b\epsilon^{\mu\nu\rho\sigma}\big(X_{a\rho}(\tau_a)-X_{b\rho}(\tau_b)\big)\frac{d^2X_{b\sigma}(\tau_b)}{d\tau_b^2}}{\big(X_a(\tau_a)-X_b(\tau_b)\big)\cdot\frac{dX_b(\tau_b)}{d\tau_b}}}\times\left\{\big(X_a(\tau_a)-X_b(\tau_b)\big)\cdot\frac{d^2X_b(\tau_b)}{d\tau_b^2}-\left(\frac{dX_b(\tau_b)}{d\tau_b}\right)^2\right\}\right]_{\tau_b=\tau^\star_{ba}}~. 
\end{split}
}
\end{equation}
We evaluate the first and second derivatives of \( X^\mu_a(\tau_a) \) and compare them with the equations of motion. These derivatives are
\begin{equation}
\label{eq:derivatives}
\begin{split}
&\frac{d X^\mu_a(\tau_a)}{d\tau_a}=\frac{p^\mu_a}{m_a}+\mathcal{O}(\lambda^{-1})~,\\& \frac{d^2 X^\mu_a(\tau_a)}{d\tau^2_a}= -C_1^{(a)\mu}\frac{1}{\tau_a^2}-\sum_{r=1}^{\infty}r(r+1)C^{(a)\mu}_{r+1}\frac{(\ln\tau_a)^r}{\tau_a^{r+2}}+\mathcal{O}(\lambda^{-3})~.
\end{split}   
\end{equation}
Using the above two equations, $F^{\mu\nu}_{(b)}(X_a(\tau_a))$ simplifies to
\begin{equation}
{\small
\begin{split}
&F^{\mu\nu}_{(b)}(X_a(\tau_a))\\&=-\frac{1}{4\pi}\frac{1}{\tau_a^2}\frac{m_a^3m_b^3}{[(p_a\cdot p_b)^2-p_a^2p_b^2]^{\frac{3}{2}}}\\&~~~~\times\left[\vphantom{m_a^2\frac{\left(p_b\cdot\left(C^{(b)}_1-C^{(a)}_1\right)\right)^2-p^2_b\left(C^{(b)}_1-C^{(a)}_1\right)^2}{(p_a\cdot p_b)^2-p^2_a p^2_b}\xi_a^2}1+2m_a\frac{p^2_bp_a\cdot\left(C^{(b)}_1-C^{(a)}_1\right)-(p_a\cdot p_b)p_b\cdot\left(C^{(b)}_1-C^{(a)}_1\right)}{(p_a\cdot p_b)^2-p^2_a p^2_b}\xi_a\right.\\&~~~~~~~~~~~~~~~~~~~~~~~~~~~~~~~~~~~\left.+m_a^2\frac{\left(p_b\cdot\left(C^{(b)}_1-C^{(a)}_1\right)\right)^2-p^2_b\left(C^{(b)}_1-C^{(a)}_1\right)^2}{(p_a\cdot p_b)^2-p^2_a p^2_b}\xi_a^2\right]^{-\frac{3}{2}}\\&~~~~\times\left[\frac{q_b(p^\mu_ap^\nu_b-p_a^\nu p_b^\mu)-g_b\epsilon^{\mu\nu\rho\sigma}p_{a\rho}p_{b\sigma}}{m_am_b}+q_b\left(C_1^{(a)\mu}-C_1^{(b)\mu}\right)\frac{p^\nu_b}{m_b}\xi_a\right.\\&~~~~~~~~~~~~~~~~~~~~~~~~~~~~~~\left.-q_b\left(C_1^{(a)\nu}-C_1^{(b)\nu}\right)\frac{p^\mu_b}{m_b}\xi_a-g_b\epsilon^{\mu\nu\rho\sigma}\left(C_{1\,\rho}^{(a)}-C_{1\,\rho}^{(b)}\right)\frac{p_{b\sigma}}{m_b}\xi_a\right]+\mathcal{O}(\lambda^{-3})~~,   
\end{split}
}
\end{equation}
and similarly $\widetilde{F}^{\mu\nu}_{(b)}(X_a(\tau_a))$ simplifies to
\begin{equation}
{\small
\begin{split}
&\widetilde{F}^{\mu\nu}_{(b)}(X_a(\tau_a))\\&=-\frac{1}{4\pi}\frac{1}{\tau_a^2}\frac{m_a^3m_b^3}{[(p_a\cdot p_b)^2-p_a^2p_b^2]^{\frac{3}{2}}}\\&~~~~\times\left[\vphantom{m_a^2\frac{\left(p_b\cdot\left(C^{(b)}_1-C^{(a)}_1\right)\right)^2-p^2_b\left(C^{(b)}_1-C^{(a)}_1\right)^2}{(p_a\cdot p_b)^2-p^2_a p^2_b}\xi_a^2}1+2m_a\frac{p^2_bp_a\cdot\left(C^{(b)}_1-C^{(a)}_1\right)-(p_a\cdot p_b)p_b\cdot\left(C^{(b)}_1-C^{(a)}_1\right)}{(p_a\cdot p_b)^2-p^2_a p^2_b}\xi_a\right.\\&~~~~~~~~~~~~~~~~~~~~~~~~~~~~~~~~~~~\left.+m_a^2\frac{\left(p_b\cdot\left(C^{(b)}_1-C^{(a)}_1\right)\right)^2-p^2_b\left(C^{(b)}_1-C^{(a)}_1\right)^2}{(p_a\cdot p_b)^2-p^2_a p^2_b}\xi_a^2\right]^{-\frac{3}{2}}\\&~~~~\times\left[\frac{g_b(p^\mu_ap^\nu_b-p_a^\nu p_b^\mu)+q_b\epsilon^{\mu\nu\rho\sigma}p_{a\rho}p_{b\sigma}}{m_am_b}+g_b\left(C_1^{(a)\mu}-C_1^{(b)\mu}\right)\frac{p^\nu_b}{m_b}\xi_a\right.\\&~~~~~~~~~~~~~~~~~~~~~~~~~~~~~~\left.-g_b\left(C_1^{(a)\nu}-C_1^{(b)\nu}\right)\frac{p^\mu_b}{m_b}\xi_a+q_b\epsilon^{\mu\nu\rho\sigma}\left(C_{1\,\rho}^{(a)}-C_{1\,\rho}^{(b)}\right)\frac{p_{b\sigma}}{m_b}\xi_a\right]+\mathcal{O}(\lambda^{-3})~~,  
\end{split}
}
\end{equation}
where $\xi_a\equiv\frac{\ln\tau_a}{\tau_a}$. By inserting the expression for the field strength and its dual into the Lorentz force law, we obtain
\begin{equation}\label{eq:Lorentz_force_law}
\begin{split}
&-C_1^{(a)\mu}-\sum_{r=1}^{\infty}r(r+1)C^{(a)\mu}_{r+1}\xi_a^r\\&\simeq-\sum_{b\neq a} \frac{p_b^2}{4\pi[(p_a\cdot p_b)^2-p_a^2p_b^2]^{\frac{3}{2}}}\\&~~\times\left[\vphantom{m_a^2\frac{\left(p_b\cdot\left(C^{(b)}_1-C^{(a)}_1\right)\right)^2-p^2_b\left(C^{(b)}_1-C^{(a)}_1\right)^2}{(p_a\cdot p_b)^2-p^2_a p^2_b}\xi_a^2}1+2m_a\frac{p^2_bp_a\cdot\left(C^{(b)}_1-C^{(a)}_1\right)-(p_a\cdot p_b)p_b\cdot\left(C^{(b)}_1-C^{(a)}_1\right)}{(p_a\cdot p_b)^2-p^2_a p^2_b}\xi_a\right.\\&~~~~~~~~~~~~~~~~~~~~~~~~~~~~~~~~~~~~~~~~~~~~~~\left.+m_a^2\frac{\left(p_b\cdot\left(C^{(b)}_1-C^{(a)}_1\right)\right)^2-p^2_b\left(C^{(b)}_1-C^{(a)}_1\right)^2}{(p_a\cdot p_b)^2-p^2_a p^2_b}\xi_a^2\right]^{-\frac{3}{2}} \\&~~ \times\Bigg[(q_aq_b+g_ag_b)\Big\{\big(p_a^2p_b^\mu-(p_a\cdot p_b)p^\mu_a\big)+m_a(p_a\cdot p_b)\left(C_1^{(b)\mu}-C_1^{(a)\mu}\right)\xi_a\\&~~~~~~~~~-m_ap_a\cdot\left(C_1^{(b)}-C_1^{(a)}\right)p_b^\mu\xi_a\Big\}-(q_ag_b-g_aq_b)m_a\epsilon^{\mu\nu\rho\sigma}p_{a\nu}\left(C_{1\,\rho}^{(b)}-C_{1\,\rho}^{(a)}\right)p_{b\sigma}\xi_a\Bigg]~. 
\end{split}    
\end{equation}
In our analysis, we ignore both self-force effects and dipole (or higher multipole) moments. The self-force effects are neglected in \eqref{eq:derivatives}. Each term on the right-hand side of this equation scales as $\mathcal{O}(\lambda^{-2})$. In contrast, contributions from the self-force are proportional to either the time derivative of acceleration (scales as $\mathcal{O}(\lambda^{-3})$) or the square of the acceleration (scales as $\mathcal{O}(\lambda^{-4})$). Consequently, these self-force terms are subdominant. The dipole (or higher multipole) moments introduce additional terms involving derivatives of the field strength and dual field strengths. Such terms are suppressed by extra factors of $\mathcal{O}(\lambda^{-1})$. Moreover and importantly these multipole couplings break worldline reparametrization invariance. This is because the dipole term in the current transforms nontrivially under reparametrizations $\tau_a \to \tilde\tau_a(\tau_a)$. In contrast, the monopole (point-charge) coupling respects this invariance.
Thus, retaining dipole or higher moments explicitly introduces a preferred parametrization of the worldline, violating the fundamental reparametrization symmetry of the relativistic point particle. For this reason, and because they are also subleading in the $\lambda$ expansion, we consistently neglect them in our leading-order analysis.
\vspace{0.4cm}\\
Setting \(\xi_a = 0\) in the above equation yields the coefficients \( C^{(a)}_{1} \) as given by
\begin{equation}\label{eq:trajcoeffnn_C_1}
C_1^{(a)\mu}=\sum_{b\neq a}\frac{(q_aq_b+g_ag_b)p_b^2}{4\pi[(p_a\cdot p_b)^2-p_a^2p_b^2]^{\frac{3}{2}}}\big(p_a^2p_b^\mu-(p_a\cdot p_b)p^\mu_a\big)~.    
\end{equation}
For \( r \geq 1 \), the coefficients \( C^{(a)\mu}_{r+1} \)\footnote{In deriving \eqref{eq:trajcoeffnn_C_r+1}, we have used the following identity to extract the trajectory coefficients $C^{(a)}$'s 
\begin{equation}
\big(1+Ax+Bx^2\big)^{-\frac{3}{2}}= \sum_{k=0}^\infty \sum_{s=0}^k (-1)^k \frac{(2k+1)!}{2^{2k} k! s!(k-s)!} A^s B^{k-s} x^{2k-s}~.    
\end{equation}} are given by
\begin{equation}\label{eq:trajcoeffnn_C_r+1}
\begin{split}
C^{(a)\mu}_{r+1}&=\frac{m_a^{r}}{2^{r}r(r+1)}\sum_{b\neq a}\frac{p_b^2}{4\pi\left[(p_a\cdot p_b)^2-p_a^2p_b^2\right]^{\frac{3}{2}}}\Bigg[(q_aq_b+g_ag_b)\big(p_a^2p_b^\mu-(p_a\cdot p_b)p^\mu_a\big)\\&~~~~\times\sum_{k=\lfloor\frac{r}{2}\rfloor}^{r}\frac{(-1)^k(2k+1)!}{k!(2k-r)!(r-k)!}\frac{1}{\left[(p_a\cdot p_b)^2-p_a^2p_b^2\right]^k}
\\&~~~~\times\left(\vphantom{\left(p_b\cdot C^{(a)}_1\right)^2}p^2_b \left(p_a\cdot C^{(b)}_1\right)+(p_a\cdot p_b)\left(p_b\cdot C^{(a)}_1\right)\right)^{2k-r}\\&~~~~~~~~~~~~~~~~~~~~~~~~~~~~~\times\left(\left(p_b\cdot C^{(a)}_1\right)^2-p^2_b\left(C^{(a)}_1-C^{(b)}_1\right)^2\right)^{r-k}
\\&~~~~+2\Big\{(q_aq_b+g_ag_b)\left(\vphantom{\left(p_b\cdot C^{(a)}_1\right)^2}(p_a\cdot p_b)\left(C^{(b)\mu}_1-C^{(a)\mu}_1\right)-\left(p_a\cdot C_1^{(b)}\right)p_b^\mu\right)\\&~~~~~~~~~~~~~~~~~~~~~~~~~~~~~~~~~~~~~~~~~~~-(q_ag_b-g_aq_b)\epsilon^{\mu\nu\rho\sigma}p_{a\nu}\left(C_{1\,\rho}^{(b)}-C_{1\,\rho}^{(a)}\right)p_{b\sigma}\Big\}\\&~~~~\times\sum_{k=\lfloor\frac{r-1}{2}\rfloor}^{r-1}\frac{(-1)^k(2k+1)!}{k!(2k+1-r)!(r-k-1)!}\frac{1}{\left[(p_a\cdot p_b)^2-p_a^2p_b^2\right]^k}
\\&~~~~~~~~~~~\times\left(\vphantom{\left(p_b\cdot C^{(a)}_1\right)^2}p^2_b \left(p_a\cdot C^{(b)}_1\right)+(p_a\cdot p_b)\left(p_b\cdot C^{(a)}_1\right)\right)^{2k+1-r}\\&~~~~~~~~~~~~~~~~~~~~~~~~~~~~~~~~~~~\times\left(\left(p_b\cdot C^{(a)}_1\right)^2-p^2_b\left(C^{(a)}_1-C^{(b)}_1\right)^2\right)^{r-k-1}\Bigg]~.
\end{split}  
\end{equation}   
It can also be verified that the trajectory coefficients $C^{(a)}_1$ and $C^{(a)}_{r+1}$ must satisfy the following conditions
\begin{equation}
p_a\cdot C^{(a)}_1=0~,\qquad p_a \cdot C^{(a)}_{r+1} = 0~,\quad\textmd{for all}~~r\ge 1~.    
\end{equation}
The physical motivation for orthogonality conditions comes from the kinematical properties of massive point particles in relativistic mechanics. In particular, the four-velocity \(u^\mu_a(\tau_a) \equiv \frac{dX^\mu_a(\tau_a)}{d\tau_a}\) of a massive particle is normalized according to
\begin{equation}
\left( \frac{dX_a(\tau_a)}{d\tau_a} \right)^2 = u_a \cdot u_a = -1~.
\end{equation}
Since this expression is a constant, its derivative with respect to the proper time \( \tau_a \) must be zero. This leads to the conclusion that the four-velocity must be orthogonal to the four-acceleration
\begin{equation}
\frac{d}{d\tau_a} \left[ \left( \frac{dX_a(\tau_a)}{d\tau_a} \right)^2 \right] = 0~,
\quad \implies 
\frac{dX_a(\tau_a)}{d\tau_a} \cdot \frac{d^2X_a(\tau_a)}{d\tau_a^2} = 0~.
\label{eq:orthogonality}
\end{equation}
Since the monomials \(\xi_a^{\,r}\) are linearly independent, each coefficient in the series must separately vanish-leading to the orthogonality conditions between the momentum and the trajectory coefficients
\begin{equation}
\begin{split}
p_a \cdot C^{(a)}_1 + \sum_{r=1}^{\infty} r(r+1)\, p_a \cdot C^{(a)}_{r+1}\, \xi_a^{\,r} = 0~,
\quad \Longrightarrow \quad 
\begin{cases}
p_a \cdot C^{(a)}_1 = 0~, \\
p_a \cdot C^{(a)}_{r+1} = 0 \quad \text{for all } r \geq 1~.
\end{cases}
\end{split}
\end{equation}
We can see that the trajectory coefficients $C_1^{(a)}$ and $C_{r+1}^{(a)}$ for $r\ge 1$ depend on both  $q_a q_b + g_a g_b$ and $q_a g_b - g_a q_b$. The physical trajectory remains invariant under $SO(2)$ duality rotations because these two combinations are themselves invariant under such transformations. A detailed proof of this invariance is given in a later section.
 Complex charges \(\mathcal{Q}_a\) and \(\mathcal{Q}_b\) are defined as
\begin{equation}\label{eq:complex_charges}
\begin{split}
\mathcal{Q}_a &= q_a + \mathrm{i}g_a~, \\
\mathcal{Q}_b &= q_b + \mathrm{i}g_b~,
\end{split}
\end{equation}
where \(q_a, q_b\) are the electric charges and \(g_a, g_b\) are the magnetic charges of particles \(a\) and \(b\), respectively. With this definition, the duality-invariant combinations appearing in the trajectory coefficients can be compactly expressed as  
\begin{equation}
\begin{split}
\operatorname{Re}(\mathcal{Q}_a \bar{\mathcal{Q}}_b) = q_a q_b + g_a g_b~, \qquad
\operatorname{Im}(\mathcal{Q}_a \bar{\mathcal{Q}}_b) = q_a g_b - g_a q_b~.
\end{split}
\end{equation}
Under duality rotation, the charge pair \( (q_a, g_a) \) transforms as  
\begin{equation}
 \begin{pmatrix} q_a \\ g_a \end{pmatrix}
 \;\longrightarrow\;
 \begin{pmatrix}
 \cos\theta & \sin\theta \\
 -\sin\theta & \cos\theta
 \end{pmatrix}
 \begin{pmatrix} q_a \\ g_a \end{pmatrix}~.
 \end{equation}
 Equivalently, if one introduces the complex charge \( \mathcal{Q}_a = q_a + \mathrm{i}g_a \), this transformation acts simply as a phase rotation 
 \begin{equation}
 \mathcal{Q}_a \;\longrightarrow\; \mathrm{e}^{\mathrm{i}\theta} \mathcal{Q}_a~.
 \end{equation}

\subsection{Trajectory coefficients for two-body scattering}
In the case of two-body scattering, we have the trajectory coefficient
\begin{equation}\label{eq:two_particle_C_1}
C_1^{(a)\mu}=\frac{(q_aq_b+g_ag_b)p_b^2}{4\pi[(p_a\cdot p_b)^2-p_a^2p_b^2]^{\frac{3}{2}}}\big(p_a^2p_b^\mu-(p_a\cdot p_b)p^\mu_a\big)~,\quad\quad\quad\textmd{for}~~b\neq a~.    
\end{equation}
We can complete the square for the following expression 
\begin{equation}\label{eq:complete_squre}
\begin{split}
 &\left[1+2m_a\frac{p^2_bp_a\cdot\left(C^{(b)}_1-C^{(a)}_1\right)-(p_a\cdot p_b)p_b\cdot\left(C^{(b)}_1-C^{(a)}_1\right)}{(p_a\cdot p_b)^2-p^2_a p^2_b}\xi_a\right.\\&~~~~~~~~~~~~~~~~~~~~~~~~~~~~~~~~~~~\left.+m_a^2\frac{\left(p_b\cdot\left(C^{(b)}_1-C^{(a)}_1\right)\right)^2-p^2_b\left(C^{(b)}_1-C^{(a)}_1\right)^2}{(p_a\cdot p_b)^2-p^2_a p^2_b}\xi_a^2\right]\\&=\left[1+m_a\frac{p^2_bp_a\cdot\left(C^{(b)}_1-C^{(a)}_1\right)-(p_a\cdot p_b)p_b\cdot\left(C^{(b)}_1-C^{(a)}_1\right)}{(p_a\cdot p_b)^2-p^2_a p^2_b}\xi_a\right]^2\\&=\left[1-\frac{(q_aq_b+g_ag_b)p_b^2}{4\pi[(p_a\cdot p_b)^2-p_a^2p_b^2]^{\frac{3}{2}}}[p_a\cdot(p_a+p_b)]m_a\xi_a\right]^2~.   
\end{split}    
\end{equation}
In the two-body scattering, the trajectory coefficient $C^{(a)}_1$ and $C^{(b)}_1$ both are linear combinations of $p_a$ and $p_b$ as clearly seen from (\ref{eq:two_particle_C_1}). So $C^{(b)}_1-C^{(a)}_1$ is also a linear combination of $p_a$ and $p_b$. If we contract $C^{(b)}_{1\,\rho}-C^{(a)}_{1\,\rho}$ with $\epsilon^{\mu\nu\rho\sigma}p_{a\nu}p_{b\sigma}$, then the result should vanish identically. The vanishing term $\epsilon^{\mu\nu\rho\sigma}p_{a\nu}\left(C^{(b)}_{1\,\rho}-C^{(a)}_{1\,\rho}\right)p_{b\sigma}$ simplifies (\ref{eq:Lorentz_force_law}) and we can find the trajectory coefficients $C_{r+1}^{(a)\mu}$.\vspace{0.4cm}\\
The simplified expression of $C_{r+1}^{(a)\mu}$ for $r\ge 1$ is given by
\begin{equation}\label{eq:two_particle_C_r+1}
 C_{r+1}^{(a)\mu}=\frac{m_a^r}{r}\left(\frac{(q_aq_b+g_ag_b)p_b^2}{4\pi [(p_a\cdot p_b)^2-p_a^2p_b^2]^{\frac{3}{2}}}\right)^{r+1}[p_a\cdot(p_a+p_b)]^r \big(p_a^2p_b^\mu-(p_a\cdot p_b)p^\mu_a\big)~,\quad\textmd{for}~~b\neq a~.  
\end{equation}
Here, we can see that the trajectory coefficients depend solely on the combination \( q_a q_b + g_a g_b \).

\section{Electric and magnetic waveforms}
We calculate the general electric and magnetic waveforms. With the particle trajectories determined, we can then evaluate both the electric and magnetic waveforms at the detector. The electric waveform \( A^\mu_e(x) \) is given by
\begin{equation}
A^\mu_e(x)=\sum_{a}\frac{q_a}{2\pi}\int_{0}^{\infty} d\tau_a~\delta_+\Big(-\big(x-X_a(\tau_a)\big)^2\Big)\frac{dX^\mu_a(\tau_a)}{d\tau_a}~~,
\end{equation}
and the magnetic waveform $A^\mu_m(x)$ is
\begin{equation}
A^\mu_m(x)=\sum_{a}\frac{g_a}{2\pi}\int_{0}^{\infty} d\tau_a~\delta_+\Big(-\big(x-X_a(\tau_a)\big)^2\Big)\frac{dX^\mu_a(\tau_a)}{d\tau_a}~.
\end{equation}
We express the detector’s position \( x^\mu \) as \( x\equiv (u + R, R \hat{\mathbf{n}}) = (u, \mathbf{0}) + R n \). In the limit of large \( R \) with \( u \) held fixed, the delta function constraint simplifies to the following expression 
\begin{equation}
\big(x - X_a(\tau_a)\big)^2 = 0~,\quad \Rightarrow \quad -2uR - 2R \, n \cdot X_a(\tau_a) + \mathcal{O}(R^0) = 0~.
\end{equation}
The term at leading order in \( R \) gives
\begin{equation}
- u = \frac{n \cdot p_a}{m_a} \tau_a + n \cdot C_1^{(a)} \ln \tau_a - \sum_{r=1}^{\infty} n \cdot C_{r+1}^{(a)} \left( \frac{\ln \tau_a}{\tau_a} \right)^r + \mathcal{O}(\lambda^{-1})~.
\end{equation}
We note that the first two terms on the right-hand side are of order \( \lambda \), whereas the rest are of order \( \lambda^0 \). Accordingly, we keep only the leading two terms and solve for \( \tau_a \) 
\begin{equation}
\tau_a = \tau_a^{\text{sol}}~, \quad \tau_a^{\text{sol}} \simeq -\frac{u \, m_a}{n \cdot p_a} \left( 1 + n \cdot C_1^{(a)} w \right)~, \quad w \equiv \frac{\ln u}{u}~,
\end{equation}

\begin{equation}
\Rightarrow \quad \xi_a^{\text{sol}} \equiv \frac{\ln \tau_a^{\text{sol}}}{\tau_a^{\text{sol}}} \simeq -\frac{n \cdot p_a}{m_a} w \left( 1 + n \cdot C_1^{(a)} w \right)^{-1}~.
\end{equation}
Hence, in the regime of large \( R \) and large retarded time \( u \), the delta function constraint simplifies to
\begin{equation}
\delta_+ \Big( - \big(x - X_a(\tau_a)\big)^2 \Big) \simeq \frac{1}{2R} \frac{\delta (\tau_a - \tau_a^{\text{sol}})}{\abs{n \cdot \frac{dX_a(\tau_a)}{d\tau_a}}}_{\tau_a=\tau_a^{\text{sol}}}~.
\end{equation}
We use 
\begin{equation}
\frac{dX^\mu_a(\tau_a)}{d\tau_a}\simeq\frac{p_a^\mu}{m_a}+C_1^{(a)\mu}\frac{1}{\tau_a}+\sum_{r=1}^{\infty}rC_{r+1}^{(a)\mu}\frac{(\ln\tau_a)^r}{\tau_a^{r+1}}+\mathcal{O}(\lambda^{-2})~.
\end{equation}
We note that the first term on the right-hand side is of order \(\lambda^0\), while the subsequent terms are of order \(\lambda^{-1}\). Because our goal is to compute the electric and magnetic waveforms up to the first subleading order in the \(\lambda^{-1}\) expansion, we must keep all these contributions. Consequently, the late-time electric radiation waveform is given by
\begin{equation}
\begin{split}
 &A^\mu_e(u,R\hat{\vb{n}})\\&\simeq-\frac{1}{4\pi R}\sum_{a}\frac{q_am_a}{n\cdot p_a}\int_{0}^{\infty} d\tau_a\,\delta\big(\tau_a-\tau_a^{\textmd{sol}}\big)\frac{\frac{p_a^\mu}{m_a}+C_1^{(a)\mu}\frac{1}{\tau_a}+\sum_{r=1}^{\infty}rC_{r+1}^{(a)\mu}\frac{(\ln\tau_a)^r}{\tau_a^{r+1}}+\mathcal{O}(\lambda^{-2})}{1+\frac{m_a}{n\cdot p_a}\frac{n\cdot C_1^{(a)}}{\tau_a}+\frac{m_a}{n\cdot p_a}\sum_{r=1}^{\infty}rn\cdot C_{r+1}^{(a)}\frac{(\ln\tau_a)^r}{\tau_a^{r+1}}+\mathcal{O}(\lambda^{-2})}~,   
\end{split}
\end{equation}
using the fact that \( n \cdot p_a \) is negative. 
\newpage \noindent We perform the integral over \( \tau_a \) and keep terms through order \( \lambda^{-1} \), resulting in 
\begin{equation}
\begin{split}
&A_e^\mu(u,R\hat{\vb{n}})\\&\simeq-\frac{1}{4\pi R}\sum_{a}\frac{q_ap^\mu_a}{n\cdot p_a}-\frac{1}{4\pi R}\sum_{a}\frac{1}{\tau_a^{\textmd{sol}}}\frac{q_am_a}{n\cdot p_a}\Bigg[C_1^{(a)\mu}-\frac{p_a^\mu}{n\cdot p_a}n\cdot C_1^{(a)}\\&~~~~~~~~~~~~~~~~~~~~~~~~~~~~~~~~~~+\sum_{r=1}^{\infty}rC_{r+1}^{(a)\mu}\left(\xi_a^{\text{sol}}\right)^r-\frac{p_a^\mu}{n\cdot p_a}\sum_{r=1}^{\infty}rn\cdot C_{r+1}^{(a)}\left(\xi_a^{\text{sol}}\right)^r\Bigg]+\mathcal{O}(\lambda^{-2})~.
\end{split}
\end{equation}
Inserting the solutions \(\tau_a^{\text{sol}}\) and \(\xi_a^{\text{sol}}\), we obtain the expression for the final electric waveform 
\begin{equation}\label{eq:electric_waveform}
\begin{split}
&A^\mu_e(u,R\hat{\vb{n}})\\&\simeq-\frac{1}{4\pi R}\sum_{a}\frac{q_ap^\mu_a}{n\cdot p_a}+\frac{1}{4\pi R}\sum_{a}\frac{q'_ap^{\prime\mu}_a}{n\cdot p'_a}\\&~~+\frac{1}{4\pi R}\sum_{r=1}^{\infty}(-1)^{r}\frac{(\ln u)^{r-1}}{u^r}\sum_{a}\frac{q_a}{n\cdot p_a}\Bigg[\left(n\cdot C_1^{(a)}\right)^{r-1}\left(\vphantom{\left(n\cdot C_1^{(a)}\right)^{r-1}}\left(n\cdot C^{(a)}_1\right)p_a^\mu-\left(n\cdot p_a\right) C^{(a)\mu}_1\right)\\&~~+\sum_{k=1}^{r-1}\frac{(r-1)!}{(k-1)!(r-k-1)!}\left(\vphantom{\left(n\cdot C_1^{(a)}\right)^{r-1}}\left(n\cdot C^{(a)}_{k+1}\right)p_a^\mu-\left(n\cdot p_a\right) C^{(a)\mu}_{k+1}\right)\left(\frac{n\cdot p_a}{m_a}\right)^k\left(n\cdot C_1^{(a)}\right)^{r-k-1}\Bigg]\\&~~~~~~~~~~~~~~~~~~~~~~~~~~+\mathcal{O}(\lambda^{-2})~.
\end{split}
\end{equation}
Likewise, the magnetic radiation waveform is given by
\begin{equation}\label{eq:magnetic_waveform}
\begin{split}
&A^\mu_m(u,R\hat{\vb{n}})\\&\simeq-\frac{1}{4\pi R}\sum_{a}\frac{g_ap^\mu_a}{n\cdot p_a}+\frac{1}{4\pi R}\sum_{a}\frac{g'_ap^{\prime\mu}_a}{n\cdot p'_a}\\&~~+\frac{1}{4\pi R}\sum_{r=1}^{\infty}(-1)^{r}\frac{(\ln u)^{r-1}}{u^r}\sum_{a}\frac{g_a}{n\cdot p_a}\Bigg[\left(n\cdot C_1^{(a)}\right)^{r-1}\left(\vphantom{\left(n\cdot C_1^{(a)}\right)^{r-1}}\left(n\cdot C^{(a)}_1\right)p_a^\mu-\left(n\cdot p_a\right) C^{(a)\mu}_1\right)\\&~~+\sum_{k=1}^{r-1}\frac{(r-1)!}{(k-1)!(r-k-1)!}\left(\vphantom{\left(n\cdot C_1^{(a)}\right)^{r-1}}\left(n\cdot C^{(a)}_{k+1}\right)p_a^\mu-\left(n\cdot p_a\right) C^{(a)\mu}_{k+1}\right)\left(\frac{n\cdot p_a}{m_a}\right)^k\left(n\cdot C_1^{(a)}\right)^{r-k-1}\Bigg]\\&~~~~~~~~~~~~~~~~~~~~~~~~~~+\mathcal{O}(\lambda^{-2})~.
\end{split}   
\end{equation}
Using conservation of electric and magnetic charges, we can verify \( n \cdot A_e=n\cdot A_m=0 \). The $u$ independent parts of $A_e^\mu$ and $A_m^\mu$ are called \textit{electric memory} and \textit{magnetic memory}, respectively. Their expressions are given by
\begin{equation}\label{eq:electric_magnetic_memory}
\begin{split}
&-\frac{1}{4\pi R}\sum_{a}\frac{q_ap^\mu_a}{n\cdot p_a}+\frac{1}{4\pi R}\sum_{a}\frac{q'_ap^{\prime\mu}_a}{n\cdot p'_a}\quad\textmd{(electric memory)}~,\\&-\frac{1}{4\pi R}\sum_{a}\frac{g_ap^\mu_a}{n\cdot p_a}+\frac{1}{4\pi R}\sum_{a}\frac{g'_ap^{\prime\mu}_a}{n\cdot p'_a}\quad\textmd{(magnetic memory)}~.
\end{split}
\end{equation}
We can see that the electric--magnetic duality $SO(2)$ transformation mixes the memory in a similar way as it mixes the electric and magnetic charges. For the incoming waveform, the memory terms will be absent, and we have to replace $u\to-u,~\ln u\to\ln\abs{u},~p^\mu_a\to p^{\prime\mu}_a,~q_a\to q'_a$ and $g_a\to g'_a$. In addition, we have to substitute $C_1^{(a)}\to C_1^{\prime(a)}$ and $C^{(a)}_{r+1}\to(-1)^rC_{r+1}^{\prime(a)}$.

\subsection{Transformation of electric and magnetic waveforms under electric--magnetic duality}
Under the $SO(2)$ duality rotation,
\begin{equation}
q_a \;\to\; q_a \cos\theta + g_a \sin\theta~, \qquad 
g_a \;\to\; -q_a \sin\theta + g_a \cos\theta~,
\end{equation}
(and analogously for any primed pair $(q_a', g_a')$), the electromagnetic potentials transform as
\begin{equation}
A_e^\mu \;\to\; A_e^\mu \cos\theta + A_m^\mu \sin\theta~, \qquad
A_m^\mu \;\to\; -A_e^\mu \sin\theta + A_m^\mu \cos\theta~.
\end{equation}
The invariance of the physical trajectory under this $SO(2)$ rotation follows from the fact that the relevant combinations appearing in the trajectory coefficients-namely $q_a q_b + g_a g_b$ and $q_a g_b - g_a q_b$ are themselves invariant under such rotations. A detailed proof of this invariance is provided. We package electric and magnetic charges into a two–component vector
\begin{equation}
\mathbold{Q}_a = 
\begin{pmatrix}
q_a \\ g_a
\end{pmatrix}~, 
\qquad
\mathbold{Q}_b = 
\begin{pmatrix}
q_b \\ g_b
\end{pmatrix}~.
\end{equation}
Under the $SO(2)$ duality rotation,
\begin{equation}
\mathbold{Q}_a' = R(\theta)\,\mathbold{Q}_a~, 
\qquad 
R(\theta) = 
\begin{pmatrix}
\cos\theta & \sin\theta \\
-\sin\theta & \cos\theta
\end{pmatrix}~,
\end{equation}
with $R^T R = I$ and $\det R = 1$.

\paragraph{1. Euclidean inner product.}
The complex charges \(\mathcal{Q}_a\) and \(\mathcal{Q}_b\) are  
\begin{equation}
\begin{split}
\mathcal{Q}_a &= q_a + \mathrm{i} g_a~, \\
\mathcal{Q}_b &= q_b + \mathrm{i}g_b~,
\end{split}
\end{equation}
with \(q_a, q_b\) denoting the electric charges and \(g_a, g_b\) the magnetic charges of particles \(a\) and \(b\), respectively. 
The real part of $\mathcal{Q}_a \bar{\mathcal{Q}}_b$ is the Euclidean dot product
\begin{equation}
\begin{split}
\Re(\mathcal{Q}_a \bar{\mathcal{Q}}_b)
&= q_a q_b + g_a g_b 
= \mathbold{Q}_a^{T} \mathbold{Q}_b~, \\
\mathbold{Q}_a' \cdot \mathbold{Q}_b'
&= (R\mathbold{Q}_a)^{T} (R\mathbold{Q}_b)
= \mathbold{Q}_a^{T} (R^{T} R) \mathbold{Q}_b
= \mathbold{Q}_a^{T} \mathbold{Q}_b~.
\end{split}
\end{equation}
Thus, the dot product is invariant, as expected for a Euclidean inner product under $SO(2)$ rotations.
\paragraph{2. Symplectic form.}

The imaginary part of $\mathcal{Q}_a \bar{\mathcal{Q}}_b$ can be written using the antisymmetric matrix
\begin{equation}
\begin{split}
J = 
\begin{pmatrix}
0 & 1 \\
-1 & 0
\end{pmatrix}~,
\end{split}
\end{equation}
so that
\begin{equation}
\begin{split}
\Im(\mathcal{Q}_a \bar{\mathcal{Q}}_b)
&= q_a g_b - g_a q_b
= \mathbold{Q}_a^{T} J \mathbold{Q}_b~.
\end{split}
\end{equation}
Under rotation,
\begin{equation}
\begin{split}
\mathbold{Q}_a'^{T} J \mathbold{Q}_b'
&= (R\mathbold{Q}_a)^{T} J (R\mathbold{Q}_b)
= \mathbold{Q}_a^{T} (R^{T} J R) \mathbold{Q}_b~.
\end{split}
\end{equation}
For any $R \in SO(2)$, one verifies that $R^{T} J R = J$, hence
\begin{equation}
\begin{split}
\mathbold{Q}_a'^{T} J \mathbold{Q}_b'
&= \mathbold{Q}_a^{T} J \mathbold{Q}_b~.
\end{split}
\end{equation}
So the symplectic form is invariant. Equivalently,
\begin{equation}
\begin{split}
\det \begin{pmatrix} q'_a & q'_b \\ g'_a & g'_b \end{pmatrix}
&= \det \big( [\mathbold{Q}'_a \ \mathbold{Q}'_b] \big)
= \det(R) \, \det \big( [\mathbold{Q}_a \ \mathbold{Q}_b] \big)
= \det \big( [\mathbold{Q}_a \ \mathbold{Q}_b] \big)~,
\end{split}
\end{equation}
since $\det R = 1$ for $SO(2)$ rotations.
Therefore, both the real and imaginary parts of $\mathcal{Q}_a \bar{\mathcal{Q}}_b$,
\begin{equation}
\begin{split}
\Re(\mathcal{Q}_a \bar{\mathcal{Q}}_b) = q_a q_b + g_a g_b~,
\qquad
\Im(\mathcal{Q}_a \bar{\mathcal{Q}}_b) = q_a g_b - g_a q_b~,
\end{split}
\end{equation}
are invariant under $SO(2)$ duality rotations.\vspace{0.2cm}\\Ultimately, the transformation rule for the electric and magnetic waveforms is a direct consequence of the identical $SO(2)$ action on the electric and magnetic charges. Specifically, consider a duality rotation by an angle \(\theta\)
\begin{equation}
\begin{pmatrix}
A_e^\mu \\
A_m^\mu
\end{pmatrix}
\;\longrightarrow\;
\begin{pmatrix}
\cos\theta & \sin\theta \\
-\sin\theta & \phantom{-}\cos\theta
\end{pmatrix}
\begin{pmatrix}
A_e^\mu \\
A_m^\mu
\end{pmatrix}~.
\end{equation}
Thus, the full set of soft electromagnetic waveforms respects electric--magnetic duality at every order in the late-time expansion, providing a powerful consistency check on the structure of the magnetic corrections to the classical soft photon theorems. 

\subsection{Time domain electric and magnetic waveforms for two-body scattering}
For the two-body scattering, the simplified electric waveform is given by
\begin{equation}\label{eq:two_particle_electric_waveform}
\begin{split}
&A^\mu_e(u,R\hat{\vb{n}})\simeq-\frac{1}{4\pi R}\Bigg[\frac{q_1p^\mu_1}{n\cdot p_1}+\frac{q_2p^\mu_2}{n\cdot p_2}-\frac{q'_{1}p^{\prime\mu}_{1}}{n\cdot p'_{1}}-\frac{q'_{2}p^{\prime\mu}_{2}}{n\cdot p'_{2}}\\&~~~~~~~~~~~~~~~~~-\left(\frac{q_1p_1^\mu}{n\cdot p_1}n\cdot p_2+\frac{q_2p_2^\mu}{n\cdot p_2}n\cdot p_1-q_1p^\mu_2-q_2p^\mu_1\right)\\&~~~~~~~~~~~~~~~~~~~~~~~~~~~\times\sum_{r=1}^{\infty}(-1)^{r}\frac{(\ln u)^{r-1}}{u^r}\left(\sigma^{\textmd{out}}_{12}\right)^{r}\left[n\cdot (p_1+p_2)\right]^{r-1}\Bigg]~,
\end{split}  
\end{equation}
where \(\sigma_{12}^{\text{out}/\textmd{in}}\), the outgoing or incoming cross-section is defined as
\begin{equation}\label{eq:cross_sections}
\sigma_{12}^{\textmd{out}}\equiv\frac{(q_1q_2+g_1g_2)p_1^2p_2^2}{4\pi\left[(p_1\cdot p_2)^2-p^2_1p^2_2\right]^{\frac{3}{2}}}~,\quad \sigma_{12}^{\textmd{in}}\equiv\frac{(q'_1q'_2+g'_1g'_2)p^{\prime 2}_{1}p^{\prime 2}_{2}}{4\pi\big[(p'_{1}\cdot p'_{2})^2-p^{\prime 2}_{1}p^{\prime 2}_{2}\big]^{\frac{3}{2}}}~.
\end{equation}
We now explain the above expression of the scattering cross-section. The prefactor $q_1q_2+g_1g_2$ in the cross-section is the strength of the amplitude $(12)_{\textmd{out}}\to(12)_{\textmd{out}}$ scattering. This is explained in \cite{Zwanziger:1968ams} by considering scattering of two non-relativistic dyons with charges $(q_1,g_1)$ and $(q_2,g_2)$. It generalizes the familiar electric Coulomb coupling $q_1 q_2$ by including the magnetic contribution $g_1 g_2$, thus accounting for both electric and magnetic charges. Importantly, this combination is invariant under electric--magnetic duality transformations.\vspace{0.2cm}\\Next, consider a theory of worldline QED \cite{Edwards:2022dbd}. We will focus on a particular four-point amplitude of the process $(12)_{\textmd{out}}\to(12)_{\textmd{out}}$. It can be obtained by inserting four photon-vertex operators $\varepsilon\cdot\dot{X}(\tau)\mathrm{e}^{\mathrm{i}p\cdot X(\tau)}$ where $\varepsilon_\mu$ is the polarization vector of the photon of momenta $p$. Two operators are inserted on the worldline of the particle $1$ and the other two operators are inserted on the worldline of the particle $2$. The propagator $\expval{X^\mu X^\nu}\sim \eta^{\mu\nu}$ makes the contraction of $p_1\cdot p_1$ and $p_2\cdot p_2$ (ignoring cross contractions). The path integral involves a Gaussian term like 
\begin{equation*}
\exp(-p_1^2a^2-2(p_1\cdot p_2)ab-p_2^2b^2)~.    
\end{equation*}
Integrating out $a,b$ will give us $\frac{1}{(p_1\cdot p_2)^2-p_1^2p_2^2}$. So, the full amplitude becomes
\begin{equation}
\mathcal{M}_{(12)_{\textmd{out}}\to(12)_{\textmd{out}}}\sim\frac{p_1^2p_2^2}{(p_1\cdot p_2)^2-p_1^2p_2^2}~.    
\end{equation}
We apply the \textit{optical theorem} in the process of two-body scattering. It relates the imaginary part of the amplitude $\Im{\mathcal{M}_{(12)_{\textmd{out}}\to(12)_{\textmd{out}}}}$ to the cross-section $\sigma^{\textmd{out}}_{12}$ in the following equation
\begin{equation}
\sigma_{12}^{\textmd{out}}\sim\frac{\Im{\mathcal{M}_{(12)_{\textmd{out}}\to(12)_{\textmd{out}}}}  }{\sqrt{(p_1\cdot p_2)^2-p_1^2p_2^2}}~.  
\end{equation}
The denominator factor of the above equation is the relative velocity between the particle $1$ and $2$. Combining all the results together, we get the formula of the desired cross-section.\vspace*{0.2cm}\\
Now summing over $r$ in (\ref{eq:two_particle_electric_waveform}), we have
\begin{equation}
\begin{split}
&A^\mu_e(u,R\hat{\vb{n}})\simeq-\frac{1}{4\pi R}\Bigg[\frac{q_1p^\mu_1}{n\cdot p_1}+\frac{q_2p^\mu_2}{n\cdot p_2}-\frac{q'_{1}p^{\prime\mu}_{1}}{n\cdot p'_{1}}-\frac{q'_{2}p^{\prime\mu}_{2}}{n\cdot p'_{2}}\Bigg]\\&~~~~~~~~~~~~~~~~~~~~~~~~~~~-\frac{1}{4\pi R}\sigma^{\textmd{out}}_{12}\frac{1}{u}\left(1+\sigma^{\textmd{out}}_{12}\left[n\cdot (p_1+p_2)\right]\frac{\ln u}{u}\right)^{-1}E^{\mu}_{\textmd{out}}~~,
\end{split}    
\end{equation}
where $E^{\mu}_{\textmd{out}}$ is given by
\begin{equation}
E^{\mu}_{\textmd{out}}=\left(\frac{q_1p_1^\mu}{n\cdot p_1}n\cdot p_2+\frac{q_2p_2^\mu}{n\cdot p_2}n\cdot p_1-q_1p^\mu_2-q_2p^\mu_1\right)~.    
\end{equation}
The early-time electric waveform is given by
\begin{equation}
\begin{split}
&A^\mu_e(u,R\hat{\vb{n}})\simeq\frac{1}{4\pi R}\left(\frac{q'_1p_1^{\prime\mu}}{n\cdot p'_1}n\cdot p'_2+\frac{q'_2p_2^{\prime\mu}}{n\cdot p'_2}n\cdot p'_1-q'_1p^{\prime\mu}_2-q'_2p^{\prime\mu}_1\right)\\&~~~~~~~~~~~~~~~~~~~~~~~~~~~\times\sum_{r=1}^{\infty}\frac{(\ln\abs{ u})^{r-1}}{u^r}\left(\sigma^{\textmd{in}}_{12}\right)^{r}\left[n\cdot (p'_1+p'_2)\right]^{r-1}~.
\end{split}  
\end{equation}
Summing over $r$, we have
\begin{equation}
\begin{split}
&A^\mu_e(u,R\hat{\vb{n}})\simeq\frac{1}{4\pi R}\sigma^{\textmd{in}}_{12}\frac{1}{u}\left(1-\sigma^{\textmd{in}}_{12}\left[n\cdot (p'_1+p'_2)\right]\frac{\ln \abs{u}}{u}\right)^{-1}E^{\mu}_{\textmd{in}}~,
\end{split}      
\end{equation}
where $E^{\mu}_{\textmd{in}}$ is given by
\begin{equation}
E^{\mu}_{\textmd{in}}=\left(\frac{q'_1p_1^{\prime\mu}}{n\cdot p'_1}n\cdot p'_2+\frac{q'_2p_2^{\prime\mu}}{n\cdot p'_2}n\cdot p'_1-q'_1p^{\prime\mu}_2-q'_2p^{\prime\mu}_1\right)~.    
\end{equation}
For the two-body scattering, the simplified magnetic waveform is given by
\begin{equation}\label{eq:two_particle_magnetic_waveform}
\begin{split}
&A^\mu_m(u,R\hat{\vb{n}})\simeq-\frac{1}{4\pi R}\Bigg[\frac{g_1p^\mu_1}{n\cdot p_1}+\frac{g_2p^\mu_2}{n\cdot p_2}-\frac{g'_{1}p^{\prime\mu}_{1}}{n\cdot p'_{1}}-\frac{g'_{2}p^{\prime\mu}_{2}}{n\cdot p'_{2}}\\&~~~~~~~~~~~~~~~~~-\left(\frac{g_1p_1^\mu}{n\cdot p_1}n\cdot p_2+\frac{g_2p_2^\mu}{n\cdot p_2}n\cdot p_1-g_1p^\mu_2-g_2p^\mu_1\right)\\&~~~~~~~~~~~~~~~~~~~~~~~~~~~\times\sum_{r=1}^{\infty}(-1)^{r}\frac{(\ln u)^{r-1}}{u^r}\left(\sigma^{\textmd{out}}_{12}\right)^{r}\left[n\cdot (p_1+p_2)\right]^{r-1}\Bigg]~.
\end{split}  
\end{equation}
Summing over $r$, we have
\begin{equation}
\begin{split}
&A^\mu_m(u,R\hat{\vb{n}})\simeq-\frac{1}{4\pi R}\Bigg[\frac{g_1p^\mu_1}{n\cdot p_1}+\frac{g_2p^\mu_2}{n\cdot p_2}-\frac{g'_{1}p^{\prime\mu}_{1}}{n\cdot p'_{1}}-\frac{g'_{2}p^{\prime\mu}_{2}}{n\cdot p'_{2}}\Bigg]\\&~~~~~~~~~~~~~~~~~~~~~~~~~~~-\frac{1}{4\pi R}\sigma^{\textmd{out}}_{12}\frac{1}{u}\left(1+\sigma^{\textmd{out}}_{12}\left[n\cdot (p_1+p_2)\right]\frac{\ln u}{u}\right)^{-1}M^{\mu}_{\textmd{out}}~~,
\end{split}    
\end{equation}
where $M^{\mu}_{\textmd{out}}$ is given by
\begin{equation}
M^{\mu}_{\textmd{out}}=\left(\frac{g_1p_1^\mu}{n\cdot p_1}n\cdot p_2+\frac{g_2p_2^\mu}{n\cdot p_2}n\cdot p_1-g_1p^\mu_2-g_2p^\mu_1\right)~.    
\end{equation}
The early-time magnetic waveform is given by
\begin{equation}
\begin{split}
&A^\mu_m(u,R\hat{\vb{n}})\simeq\frac{1}{4\pi R}\left(\frac{g'_1p_1^{\prime\mu}}{n\cdot p'_1}n\cdot p'_2+\frac{g'_2p_2^{\prime\mu}}{n\cdot p'_2}n\cdot p'_1-g'_1p^{\prime\mu}_2-g'_2p^{\prime\mu}_1\right)\\&~~~~~~~~~~~~~~~~~~~~~~~~~~~\times\sum_{r=1}^{\infty}\frac{(\ln\abs{ u})^{r-1}}{u^r}\left(\sigma^{\textmd{in}}_{12}\right)^{r}\left[n\cdot (p'_1+p'_2)\right]^{r-1}~.
\end{split}  
\end{equation}
Summing over $r$, we have
\begin{equation}
\begin{split}
&A^\mu_m(u,R\hat{\vb{n}})\simeq\frac{1}{4\pi R}\sigma^{\textmd{in}}_{12}\frac{1}{u}\left(1-\sigma^{\textmd{in}}_{12}\left[n\cdot (p'_1+p'_2)\right]\frac{\ln \abs{u}}{u}\right)^{-1}M^{\mu}_{\textmd{in}}~,
\end{split}      
\end{equation}
where $M^{\mu}_{\textmd{in}}$ is given by
\begin{equation}
M^{\mu}_{\textmd{in}}=\left(\frac{g'_1p_1^{\prime\mu}}{n\cdot p'_1}n\cdot p'_2+\frac{g'_2p_2^{\prime\mu}}{n\cdot p'_2}n\cdot p'_1-g'_1p^{\prime\mu}_2-g'_2p^{\prime\mu}_1\right)~.    
\end{equation}
\subsection{Frequency domain electric and magnetic waveforms for two-body scattering}
The electric waveform in frequency space is defined as
\begin{equation}\label{electric_waveform_fequency_space}
\widetilde{A}^\mu_e(\omega,R\hat{\vb{n}})\equiv\int_{-\infty}^{\infty} du\,\mathrm{e}^{\mathrm{i}\omega u}A^\mu_e(u,R\hat{\vb{n}})~.   
\end{equation}
We employ the expression for the waveform in the frequency space.
\begin{equation}\label{eq:Fourier_transform_identity}
\begin{split}
& \int_{-\infty}^{\infty} \frac{d \omega}{2 \pi} \mathrm{e}^{-\mathrm{i} \omega u} \omega^{k-1}\{\ln (\omega+\mathrm{i} \varepsilon)\}^k \simeq \begin{cases}-k!\mathrm{i}^{k-1} \frac{(\ln u)^{k-1}}{u^k} &~\text { for } u \rightarrow+\infty \\
0 &~\text { for } u\rightarrow-\infty\end{cases} \\
& \int_{-\infty}^{\infty} \frac{d \omega}{2 \pi} \mathrm{e}^{-\mathrm{i} \omega u} \omega^{k-1}\{\ln (\omega-\mathrm{i} \varepsilon)\}^k \simeq \begin{cases}0 & \text { for } u \rightarrow+\infty \\
+k!\mathrm{i}^{k-1} \frac{(\ln \abs{u})^{k-1}}{u^k} & \text { for } u \rightarrow-\infty\end{cases}
\end{split}
\end{equation}
where $\varepsilon$ is an infinitesimal positive number. \vspace{0.2cm}\\Thus, the frequency domain electric waveform \(\widetilde{A}^\mu_e(\omega, R\hat{\mathbf{n}})\) is given by
\begin{equation}
\begin{split}
\widetilde{A}^\mu_e(\omega,R\hat{\vb{n}})&\simeq -\frac{\mathrm{i}}{4\pi R}\frac{1}{\omega+\mathrm{i}\varepsilon}\left[\frac{q_1p^\mu_1}{n\cdot p_1}+\frac{q_2p^\mu_2}{n\cdot p_2}-\frac{q'_{1}p^{\prime\mu}_{1}}{n\cdot p'_{1}}-\frac{q'_{2}p^{\prime\mu}_{2}}{n\cdot p'_{2}}\right]  \\&~~~~~-\frac{\mathrm{i}}{4\pi R}E^\mu_{\textmd{out}}\sum_{r=1}^{\infty}\frac{1}{r!}\omega^{r-1} \left( \ln(\omega+\mathrm{i}\varepsilon) \right)^r \left(+\mathrm{i}\sigma^{\textmd{out}}_{12}\right)^{r}\left[n\cdot (p_1+p_2)\right]^{r-1}\\&~~~~~+\frac{\mathrm{i}}{4\pi R}E^\mu_{\textmd{in}}\sum_{r=1}^{\infty}\frac{1}{r!}\omega^{r-1} \left( \ln(\omega-\mathrm{i}\varepsilon) \right)^r \left(-\mathrm{i}\sigma^{\textmd{in}}_{12}\right)^{r}\left[n\cdot (p'_1+p'_2)\right]^{r-1}~.  
\end{split}    
\end{equation}
After summing over \( r \), the resummed electric waveform for two-particle scattering takes the form of
\begin{equation}
\begin{split}
&\widetilde{A}^\mu_e(\omega,R\hat{\vb{n}})\\&\simeq -\frac{\mathrm{i}}{4\pi R}\frac{1}{\omega+\mathrm{i}\varepsilon}\Bigg[\frac{q_1p^\mu_1}{n\cdot p_1}+\frac{q_2p^\mu_2}{n\cdot p_2}-\frac{q'_{1}p^{\prime\mu}_{1}}{n\cdot p'_{1}}-\frac{q'_{2}p^{\prime\mu}_{2}}{n\cdot p'_{2}}\\&~~~~~~~~~~~~~~~~~~~~~+\frac{(\omega+\mathrm{i}\varepsilon)^{+\mathrm{i}\omega\sigma^{\textmd{out}}_{12}n\cdot(p_1+p_2)}-1}{n\cdot(p_1+p_2)}E^\mu_{\textmd{out}}-\frac{(\omega-\mathrm{i}\varepsilon)^{-\mathrm{i}\omega\sigma^{\textmd{in}}_{12}n\cdot(p'_1+p'_2)}-1}{n\cdot(p'_1+p'_2)}E^\mu_{\textmd{in}}\Bigg]~.    
\end{split}    
\end{equation}
By invoking momentum conservation \( p^\mu_1 + p^\mu_2 = p_1^{\prime \mu} + p_2^{\prime \mu} \) and electric charge conservation \( q_1 + q_2 = q_1' + q_2' \), the above expression simplifies further to
\begin{equation}\label{eq:two_particle_electric_waveform_fequency_space}
\begin{split}
&\widetilde{A}^\mu_e(\omega,R\hat{\vb{n}})\\&\simeq -\frac{\mathrm{i}}{4\pi R}\frac{1}{\omega+\mathrm{i}\varepsilon}\Bigg[\frac{(\omega+\mathrm{i}\varepsilon)^{+\mathrm{i}\omega\sigma^{\textmd{out}}_{12}n\cdot(p_1+p_2)}}{n\cdot(p_1+p_2)}E^\mu_{\textmd{out}}-\frac{(\omega-\mathrm{i}\varepsilon)^{-\mathrm{i}\omega\sigma^{\textmd{in}}_{12}n\cdot(p'_1+p'_2)}}{n\cdot(p'_1+p'_2)}E^\mu_{\textmd{in}}\Bigg]~.      
\end{split} 
\end{equation}
The magnetic waveform in frequency space is defined as
\begin{equation}\label{megnetic_waveform_fequency_space}
\widetilde{A}^\mu_m(\omega,R\hat{\vb{n}})\equiv\int_{-\infty}^{\infty} du\,\mathrm{e}^{\mathrm{i}\omega u}A^\mu_m(u,R\hat{\vb{n}})~.   
\end{equation}
Thus, the frequency domain magnetic waveform \(\widetilde{A}^\mu_m(\omega, R\hat{\mathbf{n}})\) is given by
\begin{equation}
\begin{split}
\widetilde{A}^\mu_m(\omega,R\hat{\vb{n}})&\simeq -\frac{\mathrm{i}}{4\pi R}\frac{1}{\omega+\mathrm{i}\varepsilon}\left[\frac{g_1p^\mu_1}{n\cdot p_1}+\frac{g_2p^\mu_2}{n\cdot p_2}-\frac{g'_{1}p^{\prime\mu}_{1}}{n\cdot p'_{1}}-\frac{g'_{2}p^{\prime\mu}_{2}}{n\cdot p'_{2}}\right]  \\&~~~~~-\frac{\mathrm{i}}{4\pi R}M^\mu_{\textmd{out}}\sum_{r=1}^{\infty}\frac{1}{r!}\omega^{r-1} \left( \ln(\omega+\mathrm{i}\varepsilon) \right)^r \left(+\mathrm{i}\sigma^{\textmd{out}}_{12}\right)^{r}\left[n\cdot (p_1+p_2)\right]^{r-1}\\&~~~~~+\frac{\mathrm{i}}{4\pi R}M^\mu_{\textmd{in}}\sum_{r=1}^{\infty}\frac{1}{r!}\omega^{r-1} \left( \ln(\omega-\mathrm{i}\varepsilon) \right)^r \left(-\mathrm{i}\sigma^{\textmd{in}}_{12}\right)^{r}\left[n\cdot (p'_1+p'_2)\right]^{r-1}~.  
\end{split}    
\end{equation}
After carrying out the sum over \( r \), the resummed magnetic waveform for two-particle scattering becomes
\begin{equation}
\begin{split}
&\widetilde{A}^\mu_m(\omega,R\hat{\vb{n}})\\&\simeq -\frac{\mathrm{i}}{4\pi R}\frac{1}{\omega+\mathrm{i}\varepsilon}\Bigg[\frac{g_1p^\mu_1}{n\cdot p_1}+\frac{g_2p^\mu_2}{n\cdot p_2}-\frac{g'_{1}p^{\prime\mu}_{1}}{n\cdot p'_{1}}-\frac{g'_{2}p^{\prime\mu}_{2}}{n\cdot p'_{2}}\\&~~~~~~~~~~~~~~~~~~~~~+\frac{(\omega+\mathrm{i}\varepsilon)^{+\mathrm{i}\omega\sigma^{\textmd{out}}_{12}n\cdot(p_1+p_2)}-1}{n\cdot(p_1+p_2)}M^\mu_{\textmd{out}}-\frac{(\omega-\mathrm{i}\varepsilon)^{-\mathrm{i}\omega\sigma^{\textmd{in}}_{12}n\cdot(p'_1+p'_2)}-1}{n\cdot(p'_1+p'_2)}M^\mu_{\textmd{in}}\Bigg]~.    
\end{split}    
\end{equation}
By applying momentum conservation \( p^\mu_1 + p^\mu_2 = p_1^{\prime \mu} + p_2^{\prime \mu} \) and magnetic charge conservation \( g_1 + g_2 = g_1' + g_2' \), the expression of the magnetic waveform simplifies further to
\begin{equation}\label{eq:two_particle_magnetic_waveform_fequency_space}
\begin{split}
&\widetilde{A}^\mu_m(\omega,R\hat{\vb{n}})\\&\simeq -\frac{\mathrm{i}}{4\pi R}\frac{1}{\omega+\mathrm{i}\varepsilon}\Bigg[\frac{(\omega+\mathrm{i}\varepsilon)^{+\mathrm{i}\omega\sigma^{\textmd{out}}_{12}n\cdot(p_1+p_2)}}{n\cdot(p_1+p_2)}M^\mu_{\textmd{out}}-\frac{(\omega-\mathrm{i}\varepsilon)^{-\mathrm{i}\omega\sigma^{\textmd{in}}_{12}n\cdot(p'_1+p'_2)}}{n\cdot(p'_1+p'_2)}M^\mu_{\textmd{in}}\Bigg]~.      
\end{split} 
\end{equation}

\subsection{Transformation of electric and magnetic waveforms under electric--magnetic duality for two-body scattering}
We consider the electric--magnetic duality rotation acting on the asymptotic gauge potentials as
\begin{equation}\label{eq:duality}
\begin{pmatrix}
A_e^\mu \\
A_m^\mu
\end{pmatrix}
\;\longrightarrow\;
\begin{pmatrix}
\cos\theta & \sin\theta \\
-\sin\theta & \cos\theta
\end{pmatrix}
\begin{pmatrix}
A_e^\mu \\
A_m^\mu
\end{pmatrix}~.
\end{equation}
This is a $SO(2)$ rotation in the space of dyonic fields. To determine how the full waveforms transform, we use the explicit expressions for the retarded-time waveforms in two-body scattering with dyonic charges $(q'_i,g'_i)$ and $(q_i,g_i)$ for the in- and out-states.
\paragraph{Duality action on charges.}
For the duality \eqref{eq:duality} to be a symmetry of the theory, the dyonic charges must rotate as
\begin{equation}\label{eq:charge_rot}
\begin{pmatrix} q_i \\ g_i \end{pmatrix}
\rightarrow
\begin{pmatrix}
\cos\theta & \sin\theta \\
-\sin\theta & \cos\theta
\end{pmatrix}
\begin{pmatrix} q_i \\ g_i \end{pmatrix}~,
\qquad
\begin{pmatrix} q'_i \\ g'_i \end{pmatrix}
\rightarrow
\begin{pmatrix}
\cos\theta & \sin\theta \\
-\sin\theta & \cos\theta
\end{pmatrix}
\begin{pmatrix} q'_i \\ g'_i \end{pmatrix}\quad\textmd{for}~~i=1,2~.
\end{equation}
Under this transformation, the combinations appearing in the waveforms mix as
\begin{equation}
E^\mu_{\out/\inm} \rightarrow \cos\theta\, E^\mu_{\out/\inm} + \sin\theta\, M^\mu_{\out/\inm}~,
\qquad
M^\mu_{\out/\inm} \rightarrow -\sin\theta\, E^\mu_{\out/\inm} + \cos\theta\, M^\mu_{\out/\inm}~.
\end{equation}
Crucially, the scalar quantity
\begin{equation}
q_1 q_2 + g_1 g_2 = \mathbold{Q}_1 \cdot \mathbold{Q}_2~,
\qquad
\mathbold{Q}_i = \begin{pmatrix} q_i \\ g_i \end{pmatrix}\quad\textmd{for}~~i=1,2
\end{equation}
is invariant under $SO(2)$ rotations, and hence $\sigma_{12}^{\out/\inm}$ are unchanged.

\paragraph{Transformed waveforms in position space.}

Using the linearity of the waveform expressions in $q_i$ and $g_i$, and the charge rotation \eqref{eq:charge_rot}, the duality-rotated waveforms are
\begin{align}
A_e^{\mu}(u,R\hat{\vb{n}}) &\rightarrow \cos\theta\, A_e^\mu(u,R\hat{\vb{n}}) + \sin\theta\, A_m^\mu(u,R\hat{\vb{n}})~, \label{eq:Ae_rot} \\
A_m^{\mu}(u,R\hat{\vb{n}}) &\rightarrow -\sin\theta\, A_e^\mu(u,R\hat{\vb{n}}) + \cos\theta\, A_m^\mu(u,R\hat{\vb{n}})~. \label{eq:Am_rot}
\end{align}
This follows directly from substituting the rotated charges into the definitions of $E^\mu_{\out/\inm}$ and $M^\mu_{\out/\inm}$ and noting the invariance of $\sigma_{12}^{\out/\inm}$ and kinematic factors.

\paragraph{Duality in frequency space.}
Because the Fourier transform is linear, the duality acts identically in frequency space
\begin{align}
\widetilde{A}_e^{\mu}(\omega,R\hat{\vb{n}}) &\rightarrow \cos\theta\, \widetilde{A}_e^\mu(\omega,R\hat{\vb{n}}) + \sin\theta\, \widetilde{A}_m^\mu(\omega,R\hat{\vb{n}})~, \label{eq:Ae_omega_rot} \\
\widetilde{A}_m^{\mu}(\omega,R\hat{\vb{n}}) &\rightarrow -\sin\theta\, \widetilde{A}_e^\mu(\omega,R\hat{\vb{n}}) + \cos\theta\, \widetilde{A}_m^\mu(\omega,R\hat{\vb{n}})~. \label{eq:Am_omega_rot}
\end{align}
Equivalently, one may rotate the charges using \eqref{eq:charge_rot}, which yields the same result due to the linearity of $E^\mu_{\out/\inm}$ and $M^\mu_{\out/\inm}$ in the charges and the invariance of $\sigma_{12}^{\out/\inm}$.\vspace{0.2cm}\\The electric and magnetic waveforms transform as a doublet under the $SO(2)$ electric--magnetic duality
and analogously in frequency space. This transformation is consistent with the rotation of dyonic charges and leaves the scattering cross-sections $\sigma_{12}^{\out/\inm}$ invariant, ensuring the covariance of the full waveform structure.

\section{Electric and magnetic waveforms in the probe approximation}
We consider a scattering process involving $N$ light probe particles, each characterized by an electric charge $q_a$, a magnetic charge $g_a$, a mass $m_a$, and outgoing momentum $p_a$, for $a = 1, 2, \dots, N$. These probes interact with a heavy dyonic scatterer of mass $M$, which carries an electric charge $Q$ and a magnetic charge $G$. The heavy scatterer is taken to be at rest and acts as a static source. The probe limit is defined by the conditions
\begin{equation}\label{eq:probelimit}
    M \gg m_a~,\quad Q \gg q_a~,\quad G \gg g_a\quad\text{and}\quad \frac{Q}{M} \ll \frac{q_a}{m_a}~,~~\frac{G}{M} \ll \frac{g_a}{m_a}~, \quad \forall a = 1, 2, \dots, N~.
\end{equation}
The additional constraint $\frac{Q}{M} \ll \frac{q_a}{m_a}$ and $\frac{G}{M} \ll \frac{g_a}{m_a}$ ensures that the heavy scatterer remains effectively stationary under the electromagnetic influence of the light probes, and its radiative effects may be neglected.
Physically, this limit ensures that the heavy dyon acts as a fixed classical background field. Its enormous inertia ($M\gg m_a$) and small charge-to-mass ratios ($Q/M,\, G/M$) imply that the backreaction of the probes on the heavy particle’s trajectory and radiation is negligible. The heavy object sources a long-range electromagnetic field that shapes the motion of the light probes but is itself unaffected by them. 
The four-momenta of the heavy scatterer and the probe particles are parameterized as follows
\begin{equation}
    P = \left(M,\boldsymbol{0}\right)~, \qquad
    p_a = \frac{m_a}{\sqrt{1 - \beta_a^2}} \left(1, \boldsymbol{\beta}_a\right)~, \quad \forall a = 1, 2, \dots, N~.
\end{equation}
Under this probe limit, the trajectory coefficients for the probe particles receive contributions exclusively from the long-range electromagnetic field generated by the heavy scatterer.
\bigskip
The leading-order trajectory correction $C_1^{(a)}$ simplifies to
\begin{equation}
    C_1^{(a)} \simeq -\frac{\big(q_a Q+g_a G\big)}{4\pi} \frac{\sqrt{1 - \beta_a^2}}{m_a \beta_a} \left(1, \frac{\boldsymbol{\beta}_a}{\beta_a^2}\right)~, \quad \forall a = 1, 2, \dots, N~.
\end{equation}
We get
\begin{equation}
    C_{r+1}^{(a)} = -\frac{1}{r} \frac{\big(q_a Q+g_a G\big)}{4\pi} \frac{\sqrt{1 - \beta_a^2}}{m_a \beta_a} \left( \frac{(q_a Q+g_a G)}{4\pi} \frac{(1 - \beta_a^2)}{m_a \beta_a^3} \right)^r \times \left(1, \frac{\boldsymbol{\beta}_a}{\beta_a^2}\right)~.
\end{equation}
Now, we examine the spatial component of the late-time multi-particle waveform expression under the probe approximation. By substituting the trajectory coefficients, we obtain the spatial asymptotic electric waveform 
\begin{equation}\label{eq:probe_electric_wavefrom}
\begin{split}
    A^i_e(u, R \hat{\bf{n}}) &\simeq \frac{1}{4\pi R} \sum_{a=1}^{N} q_a \left( \frac{\beta_a^i}{1 - \hat{\bf{n}} \cdot \boldsymbol{\beta}_a} - \frac{\beta_a^{\prime i}}{1 - \hat{\bf{n}} \cdot \boldsymbol{\beta}_a^{\,\prime}} \right) \\
    &~~~~- \frac{1}{4\pi R} \sum_{r=1}^{\infty} \frac{(\ln u)^{r-1}}{u^r} \sum_{a=1}^{N} \left( \frac{(Q q_a+Gg_a)}{4\pi} \frac{(1 - \beta_a^2)^{\frac{3}{2}}}{m_a \beta_a^3} \right)^r \frac{q_a \beta_a^i}{1 - \hat{\bf{n}} \cdot \boldsymbol{\beta}_a}~,
\end{split}
\end{equation}
where $\{ \boldsymbol{\beta}_a^{\,\prime} \}$ denote the velocities of the incoming probe particles. 
The spatial asymptotic magnetic waveform is given by 
\begin{equation}\label{eq:probe_magnetic_wavefrom}
\begin{split}
    A^i_m(u, R \hat{\bf{n}}) &\simeq \frac{1}{4\pi R} \sum_{a=1}^{N} g_a \left( \frac{\beta_a^i}{1 - \hat{\bf{n}} \cdot \boldsymbol{\beta}_a} - \frac{\beta_a^{\prime i}}{1 - \hat{\bf{n}} \cdot \boldsymbol{\beta}_a^{\,\prime}} \right) \\
    &~~~~- \frac{1}{4\pi R} \sum_{r=1}^{\infty} \frac{(\ln u)^{r-1}}{u^r} \sum_{a=1}^{N} \left( \frac{(Q q_a+G g_a)}{4\pi} \frac{(1 - \beta_a^2)^{\frac{3}{2}}}{m_a \beta_a^3} \right)^r \frac{g_a \beta_a^i}{1 - \hat{\bf{n}} \cdot \boldsymbol{\beta}_a}~.
\end{split}
\end{equation}
We can evaluate the temporal components of the electric and magnetic potentials as $A_e^0=\hat{\vb{n}}\cdot \vb{A}_e$ and $A_m^0=\hat{\vb{n}}\cdot \vb{A}_m$. For the incoming electric and magnetic waveforms, $u$ will be negative and the memory term will be absent in the waveform. We have to do the following substitutions $u\to-u,~\ln u\to\ln\abs{u}$ and $\boldsymbol{\beta}_a\to\boldsymbol{\beta}'_a$.

\section{Conclusions and future directions}

In this work, we have provided a complete and systematic derivation of the all-orders classical soft photon theorem in the presence of both electric and magnetic charges. By employing a local, dual-symmetric formulation of electrodynamics with two independent gauge potentials, we have calculated the full tower of subleading corrections to the electromagnetic waveform emitted during a generic dyonic scattering process. Our results extend the known universal behavior of radiation at asymptotic early and late retarded times \(u\) to include terms of the form \(u^{-r-1}(\ln|u|)^r\) for all \(r \geq 0\), with explicit expressions given solely in terms of the asymptotic kinematic data (masses and momenta) and the electric and magnetic charges of the scattered particles. The entire soft expansion, including all magnetic corrections, has been constructed to be manifestly covariant under $SO(2)$ electric--magnetic duality rotations. This provides a powerful and non-trivial consistency check of our formalism, as the physical trajectories and the final waveforms transform correctly under the duality, with all dependence on charges entering through the duality-invariant combinations \(\text{Re}(\mathcal{Q}_a\bar{\mathcal{Q}}_b) = q_a q_b + g_a g_b\) and \(\text{Im}(\mathcal{Q}_a\bar{\mathcal{Q}}_b) = q_a g_b - g_a q_b\). We derived explicit, recursive formulas for the trajectory coefficients \(C_{r+1}^{(a)\mu}\) that govern the late-time behavior of the scattered dyons. For the specific and physically relevant case of two-body scattering, these relations simplify, allowing us to obtain resummed, closed-form expressions for the complete low-frequency electric and magnetic waveforms in both the time and frequency domains. Building on the connection between classical soft expansions and quantum soft theorems, we conjecture that the quantum soft factor for the emission of a single soft photon-defined as the ratio of the $\mathcal{S}$-matrix in Quantum Electro-Magneto Dynamics (QEMD), which includes magnetic monopoles, with one additional external soft photon to the $\mathcal{S}$-matrix without that soft photon can be written as
\begin{equation}\label{eq:Smatrix}
\mathbb{S}^{\text{QEMD}}_{em}=\mathrm{i}(4\pi R) \Big(\varepsilon_{e\,\mu}(p)\widetilde{A}^\mu_e(\omega,R\hat{\vb{n}})+\varepsilon_{m\,\mu}(p)\widetilde{A}^\mu_m(\omega,R\hat{\vb{n}})\Big)~,\quad\textmd{where}~~p\equiv\omega(1,\hat{\vb{n}})  
\end{equation}
where $\varepsilon_e(p)$ and $\varepsilon_m(p)$ are the polarization vectors in the electric and magnetic sectors, respectively. In conclusion, this work completes the classical picture of soft radiation in electromagnetic scattering by fully incorporating magnetic monopoles and deriving the consequent corrections to all orders. 

\subsection*{Future directions.}

\paragraph{Connection to asymptotic symmetries and generalized charges.} 

Understanding the link between classical electromagnetic waveforms and asymptotic symmetries is essential. In particular, it has been shown in \cite{Campiglia:2019wxe, AtulBhatkar:2019vcb, Agrawal:2023zea, Choi:2024ygx, Choi:2024mac, Choi:2024ajz, Briceno:2025ivl, Boschetti:2025tru}-that the coefficient of the \(1/u\) term in both electromagnetic and gravitational radiation waveforms at early and late retarded times is directly tied to asymptotic symmetries and their corresponding conserved charges. The established link between asymptotic symmetries and the leading soft behavior suggests that the entire tower of subleading terms in the soft expansion-scaling as \(u^{-r-1}(\ln u)^r\) for \(r \geq 0\) may originate from an extended algebra of asymptotic symmetries, with each term governed by a corresponding generalized conserved charge. If so, such a structure would suggest the existence of an infinite hierarchy of generalized conserved charges at null infinity, each governing a specific logarithmic order in the late-time waveform. Uncovering such a symmetry principle would not only provide a profound conceptual underpinning for the all-orders classical soft theorems-including the magnetic corrections presented in this work-but could also offer new insights into the infrared structure of gauge theories with dyonic sources.

\paragraph{All-orders soft expansion and celestial amplitudes.}
A natural direction for future work is to reinterpret our all-orders classical soft photon expansion within the framework of celestial amplitudes, where four-dimensional scattering amplitudes are mapped to two-dimensional correlators on the celestial sphere via Mellin transforms. In this language, each term in our all-orders soft expansion should correspond to a distinct celestial operator, producing an infinite tower of soft constraints that extends far beyond the familiar leading and subleading soft theorems. The manifest electric--magnetic duality covariance of our expansion suggests the existence of a complexified, infinite-dimensional celestial current algebra relevant for theories with dyons. The associated analytic structures may provide essential data for a complete celestial reconstruction of classical electromagnetic radiation.

\paragraph{Carrollian amplitudes and null holography.}
A complementary direction is to reformulate our all-orders soft expansion in the setting of Carrollian amplitudes, which describe radiative data as fields living on the null boundary $\mathscr{I}^\pm$. Under Fourier transform, each inverse-power and logarithmic term in retarded time $u$ maps to a Carrollian operator with definite scaling, generating an infinite ladder of soft modes analogous to the celestial tower. The duality-covariant structure of our soft solution further hints at a complex Carrollian current algebra at null infinity, potentially enlarging known Carrollian symmetries. This perspective may enable a full Carrollian reconstruction of classical electromagnetic radiation, particularly in the presence of magnetic charges, and thus offers a promising avenue for future investigation.

\paragraph{Inclusion of spin and higher multipole moments.} A crucial generalization is to incorporate the classical spin and intrinsic electromagnetic dipole (and higher multipole) moments of the dyons.

\paragraph{Worldsheet formulation of classical soft theorems.} A particularly profound direction is to recast the classical soft theorems, including the magnetic corrections presented here, within a worldsheet formalism. 

\subsection*{Formal QFT connections.}

\begin{itemize}
    \item \textbf{From classical waveforms to quantum loops.} A natural next step is to verify, through explicit loop calculations in a low-energy effective QFT of dyons, that the classical limit of the quantum soft factors reproduces our all-orders waveforms, firmly establishing the quantum-classical correspondence.

     \item \textbf{Infrared finite observables and dressed states.} It would be interesting to explore the implications of these results for constructing infrared-finite scattering amplitudes and asymptotic states (``dressed states'') in theories with dyons, where the classical waveforms should be related to the coherent states of soft photons dressing the asymptotic states.

\end{itemize}

\paragraph{Duality groups and Langlands dual.}
Montonen--Olive/GNO duality suggests that under S-duality, a gauge theory with gauge group \( G \) and complexified coupling \( \tau \) is mapped to a dual theory whose gauge group is the Langlands (GNO) dual \( {}^{L}\!G \), and whose electric and magnetic charges are exchanged by the corresponding lattice duality. In \(\mathcal{N}=4\) supersymmetric Yang--Mills theory, S-duality is believed to be an exact symmetry and acts on the coupling as an \( SL(2,\mathbb{Z}) \) modular transformation,
\[
\tau \;\longrightarrow\; \frac{a\tau + b}{c\tau + d}~, \qquad 
\begin{pmatrix} a & b \\ c & d \end{pmatrix} \in SL(2,\mathbb{Z})~.
\]
The complex coupling is defined as
\[
\tau = \frac{\theta}{2\pi} + \frac{4\pi\mathrm{i}}{g_{\text{YM}}^2}~.
\]
Any soft expansion that is duality-covariant must therefore transform consistently under this action, such that the electric and magnetic sectors are mixed according to the modular transformation properties dictated by the duality group.

\subsection*{Acknowledgements}
The research of SD is supported by the Shuimu Tsinghua Scholar Program of Tsinghua University and the Beijing Natural Science Foundation of China (Grant No. IS25035, Project Code: 04151000525). The research of PR is supported by the Department of Atomic Energy, Government of India.




\bibliography{References}
\bibliographystyle{JHEP}

\end{document}